\documentstyle[aps]{revtex}

\input{epsf}
\begin{document}
\title{{\LARGE  $(Sr/Ca)_{14}Cu_{24}O_{41}$ spin ladders studied by
NMR under pressure.}}
\date{\today}
\author{{\it {\Large Y.Piskunov$^{1}$, D.J\'{e}rome$^{1}$, P.Auban-Senzier$^{1}$,
P.Wzietek$^{1}$,C.Bourbonnais$^{2}$, U.Ammerhal$^{3,4}$, G.Dhalenne$^{3}$
and A.Revcolevschi$^{3}$}}}
\address{\vspace*{5mm} $^{1}${\it Laboratoire de Physique des Solides (UMR 8502)}\\
Universit\'{e} Paris-Sud, 91405, Orsay, France\\
$^{2}${\it D\'epartement de Physique, Universit\'e de} Sherbrooke\\
 Sherbrooke, Qu\'ebec,J1K2R1,Canada\\
$^{3}${\it Laboratoire de Chimie des Solides (UMR 8648)\\ Universit\'{e}}
Paris-Sud, 91405, Orsay, France\\
$^{4}${\it II. Physikalisches Institut, Universit\"at zu K\"oln\\ Z\"ulpicher}
Str. 77, 50937 K\"oln, Germany}
\date{\today}
\maketitle
\pacs{AAA}

\begin{abstract}
$^{63}$Cu-NMR  measurements have been performed on two-leg
hole-doped spin ladders $Sr_{14-x}Ca_{x}Cu_{24}O_{41} $ single crystals $
0\leq x\leq 12$ at several pressures up to the pressure domain 
 where the stabilization of a superconducting ground state
can be achieved. The data reveal a marked decrease of the spin gap derived from
Knight shift measurements upon
$Ca$ substitution and also under pressure and confirm the onset of low lying spin
excitations around $P_{c}$ as previously reported. The spin gap in $
Sr_{2}Ca_{12}Cu_{24}O_{41}$ is strongly reduced above $20 kbar$. However, the
data of an experiment performed at $P=36 kbar$ where superconductivity has been detected at $6.7K$ by an
inductive technique    have shown that  a significant amount of  spin excitations  remains gapped at 80K
 when superconductivity sets in. The standard relaxation model
with two and three-magnon modes  explains fairly
well the activated relaxation data in the intermediate temperature regime corresponding
to gapped spin excitations using the  spin gap data derived from Knight shift
experiments.The data of Gaussian relaxation rates of heavily doped samples support the limitation of the
coherence lenght at low temperature by the average distance between doped holes. We discuss the interplay
between superconductivity and the spin gap and suggest that these new results support the exciting prospect
of superconductivity induced by the interladder tunnelling of preformed pairs as long as the pressure
remains lower than the pressure corresponding to the maximum of the superconducting critical temperature. 
\end{abstract}

\smallskip

\section{Introduction}

The discovery of a superconducting state under pressure in a hole-doped
spin-ladder cuprate $Sr_{14-x}Ca_{x}Cu_{24}O_{41}$ with $x=13.6$ (sintered
powder) \cite{Uehara96} and $x=11.5$ (single crystal) \cite{Nagata97}, has
renewed the interest for superconductivity in low dimensional systems and
its interplay with magnetism. These systems can be viewed as a link between
high $T_{c}$ cuprates and quasi one dimensional organics which are two
classes of exotic superconductors where the mechanism of superconductivity
is still highly debated \cite{Dagotto96,Bourbonnais99}.

\bigskip

Undoped spin-ladders of general formula $Sr_{n}Cu_{n+1}O_{2n+1}$, where the
copper valence is exactly $2$ ($Cu^{2+}$ with spin $S=1/2$), consist of
cuprate ladder planes separated by $Sr$ layers \cite{McCarron88,Siegrist88}.
Their magnetic properties depend on the number of legs in the ladders \cite
{Dagotto96}. In case of an even number of legs the formation of spin
singlets on each rung leads to the opening of a spin gap in the spin
excitations. In the two-leg ladder compound $SrCu_{2}O_{3}$ the spin gap can
be seen by the exponential drop of the spin susceptibility \cite{Azuma94}.
This singlet ground state is not possible for an odd number of legs and a
finite value of the spin susceptibility is reached at low temperature for
instance in the three-leg compound $Sr_{2}Cu_{3}O_{5}$ \cite{Azuma94}.

\bigskip

The $Sr_{14-x}Ca_{x}Cu_{24}O_{41+\delta }$ series has a more complicated
structure built from the piling up of $CuO_{2}$ chains, $Sr$ (or $Ca$) and $%
Cu_{2}O_{3}$ two-leg ladders layers with an orthorhombic symmetry \cite
{Roth89}. The interlayer distance (along the $b$ axis), which is $1.6\AA $
in $x=0$ compound, is shortened upon $Ca$ substitution. Moreover, there is a
misfit between the lattice parameters of chains and ladders sublattices ($%
10\times c_{chain}=7\times c_{ladder}$). This system is inherently hole
doped as the stoechiometry implies an average copper valence of $2.25$. However, optical 
measurements \cite{Osafune97} have shown that holes for the 
$x=0$ compound are residing mainly in the chains leading in turn to an insulating
behavior and to a magnetic ordering revealed by neutron scattering experiments
\cite{Regnault99} with a valency of
$ 2.06$ for the $Cu$ ladder sites. Then, although $Ca$ and
$Sr$ are isovalent atoms, holes are transfered from the chains to the ladders upon $Ca$
substitution. Consequently, the copper valency in the ladders increases from 
$2.06$ up to $2.20$ ($0.2$ hole per $Cu$) as $x$ increases from $0$ to $11$
\cite{Osafune97}. At the same time, the longitudinal conductivity (along the 
$c$ axis) increases, leading to quasi one dimensional ($Q-1-D$) metallic properties
\cite{Motoyama97,Nagata98}. It has been suggested that holes doped into the ladders will
share  common rungs since it is the configuration which minimizes the number of damaged spin
singlets, \cite{Dagotto92}. As far as the transverse a-axis is concerned, a confinement of
holes pairs on the ladders  below a confinement temperature called
$T^{\star }$ which is above room temperature for all samples with $x$ $\leq 8$ and drops to
below $ 300$K for $x=12$. However, as for undoped ladder compounds, a spin gap is
still observed in the ladder plane and a smaller spin gap $(\approx 140K)$ is also
present in the chain sublattice. The existence of a spin gap on ladders was demonstrated
by
$^{63}Cu$-NMR measurements which can easily separate the contributions coming from
copper nuclei belonging to chains and ladders. The spin part of the NMR shift, which is
directly proportional to the spin susceptibility, allows an accurate determination of
these gaps. Upon $Ca$ substitution, the amplitude of the spin gap in the
ladders decreases from $500K$ \cite{Kumagai97,Magishi98} or $430K$ \cite
{Takigawa98} for $x=0$ down to $250K$ for $x$=12 \cite{Magishi98,Takigawa98}
while it remains practically unchanged in the chains ($125-140K$) \cite
{Kumagai97,Takigawa98}. Therefore, the spin gapped structure survives the
existence of a finite concentration of holes in the ladders in agreement
with  theoretical suggestions \cite{Dagotto92,Noack94,Poilblanc94}. For
samples with the largest $Ca$ concentration, an applied pressure in the
range $30-80kbar$ stabilizes a superconducting ground state in the ladder
planes with a transition temperature passing through a maximum at $10K$
around $40kbar$ \cite{Nagata98}.

\bigskip

The great interest of superconducting spin ladders lies in the theoretical
prediction of superconductivity in the $d$-wave channel which is a direct
consequence of the spin gapped character of the magnetic excitations \cite
{Dagotto92}. The possible link between the predicted superconducting phase
of isolated ladders and the ($SC$) phase stabilized under high pressure in $%
(Sr/Ca)_{14}Cu_{24}O_{41}$ is therefore a challenge in the physics of
strongly correlated low dimensional fermions.

\bigskip

An early pressure study carried out on $Sr_{2}Ca_{12}Cu_{24}O_{41}$\cite
{Mayaffre98} concluded to the existence of low lying spin excitations
under high pressure conditions {\it i.e.} $P=30-32kbar$ when
superconductivity appears at $T_{c}\approx 5$K. However, recent NMR studies
at $17kbar$ and inelastic neutron scattering experiments at $21 kbar$ argued that 
the spin gap  does not change under pressure 
\cite{Mito99,Katano199,Katano299}. Consequently, it is a very important
matter to study how the transient domain between the spin gap regime and the
situation with low lying spin excitations develops under pressure.

\bigskip

The compound $La_{5}Ca_{9}Cu_{24}O_{41}$ also investigated in this work is
the closest to the situation of the undoped spin ladders although displaying
the chains and ladders structure similar to the $(Sr/Ca)_{14}Cu_{24}O_{41}$
series with an average $Cu$ valence of $2.04$ as compared to $2.25$ in the
latter series \cite{Udo,Regnault99}.

\bigskip

Our work presents the results of a study of the whole $
Sr_{14-x}Ca_{x}Cu_{24}O_{41}$ series (called $Cax$ from now) on including the parent compound $
La_{5}Ca_{9}Cu_{24}O_{41}$ \cite{Udo} ($La5$) via NMR shifts and relaxation measurements under
pressure up to $ 36 kbar$ which is a pressure well above the critical pressure
($\approx 30 kbar$) needed for the stabilization of superconductivity in $
Sr_{2}Ca_{12}Cu_{24}O_{41}$. Some preliminary measurements performed on $Ca12$ only
 up to $32 kbar$  have appeared in reference
\cite{Piskunov00}.
 Section II provides some experimental details. NMR  shifts and relaxation 
results are presented in sec.III. These results are discussed in sec.IV, in terms of
magnon excitations in undoped and doped spin ladders. We shall see how this picture breaks down when the
superconducting phase is approached. The interpretation of the Gaussian relaxation time in
sec. V provides an access to the spin correlation lenght. Finally, the relation between  spin gap and
pressure-induced superconductivity is discussed in the sec. VI.

\section{Experimental}

\bigskip

Experiments have been carried out on single crystalline samples of $
Sr_{14-x}Ca_{x}Cu_{24}O_{41}$ $(x=0,2,5,8,9,12)$ and $
La_{5}Ca_{9}Cu_{24}O_{41}$ grown by the traveling solvent floating zone
method \cite{Regnault99}. $^{63}Cu(I=3/2)$ NMR measurements have been
performed at a fixed magnetic field of $9.3$ Tesla ($B\parallel b$) {\it via} a Fourier
transform of the spin echo on the central transition $(1/2,-1/2)$ using
phase alternation techniques. The echo signal was obtained after 
the pulse sequence $(\tau _{p})_{x}-t_{del}-(\tau _{p})_{y}-t-echo$ with
$\tau _{p}\leq 1.5$ $\mu s$. The $Q$-factor of the resonance circuit did not exceed
 $50$. This provides a rather uniform irradiation of the central line for the
measurements down to $80$ K. The broad $^{63}Cu$ spectrum at low temperature
was obtained by summing up the Fourier transforms of spin echo signals taken
at different frequencies with a frequency increment such as $\Delta \nu
_{in}\sim 100\div 150$ kHz which is kept smaller than the full-width at
half-maximum (FWHM) of a single Fourier spectrum. The possible influence of the phase adjustment on
the line shape  for each Fourier transformed
subspectra was minimized in the following way. The resonance circuit was tuned
at every  step $\Delta \nu _{in}$. A subspectrum was obtained at a fixed
phase and then was passed through a rectangular band-pass digital filter $
(-1.7\Delta \nu _{in};+1.7\Delta \nu _{in})$. The comparison between different spectra
 $\sim (U_{\chi ^{\prime }}^{2}+U_{\chi ^{\prime \prime }}^{2})^{1/2}
$ shows that the peak position  can be measured with an accuracy of $\pm 25$
kHz.

\bigskip 

NMR data have been obtained in a seemingly non magnetic high hydrostatic
pressure cell. However, in order to take into account the unavoidable slight
shift of the external field generated by the tungsten carbide piston close
to the sample, the NMR signal from the copper RF coil has been used as a
field marker (known to be only weakly pressure dependent)\cite{Benedek58}.
The influence of the piston was manifested in an additional $T$-independent
positive shift of the resonance frequency for $^{63}Cu$ in the $RF$ coil, $
^{63}\nu (metal)$, remaining within $\Delta \nu _{add}\sim 120\div 170 kHz$
depending on pressure as well as in a slight broadening of the $^{63}Cu(metal)$
NMR line from $20 kHz$ (FWHF) (without piston) to $30 kHz$ in the
presence of the piston. The FWHM of $^{63}Cu(metal)$ spectra never exceeded $50
 kHz$ at any pressure and temperature.Consequently, we can be confident that all shifts
or line broadenings reported in this study exceeding $50 kHz$ are related to intrinsic
properties of the samples.  Shift values reported in this work have been
calculated relative to the
$^{63}Cu$ resonance in a diamagnetic substance, namely $^{63}\nu _{0}=^{63}\nu
(metal)/1.0023$, where 
$^{63}\nu (metal)$ is the resonance frequency of copper metal nuclei.The magnetic
shift of the $^{63}Cu-$NMR line was determined using a simulation software taking into
account quadrupolar corrections to the Zeeman frequency in a second order perturbation
theory.

\bigskip  

For a magnetic field $B$ parallel to the $\alpha$-axis of the sample
 the resonance frequency of the central transition reads\cite
{Baugner69,Creel74},

\begin{equation}
\nu _{\left( 1/2,-1/2\right) ,\alpha}=(1+K_{\alpha})\nu _{0}+\frac{\left( \nu
_{\beta}-\nu _{\gamma}\right) ^{2}}{12\nu _{0}\left( 1+K_{\alpha}\right) }
\end{equation}
where $K_{\alpha}$, $\nu _{0}$, $\nu _{\beta,\gamma}$ are the NMR shift, the Larmor
frequency in a diamagnetic substance and the quadrupolar tensor components
respectively. 

For a magnetic field $B$ parallel to the $b$-axis of the sample ({\it i.e.}
the transverse axis corresponding to the alternate packing axis of chains
and ladders) the resonance frequency of the central transition reads,

\begin{equation}
\nu _{\left( 1/2,-1/2\right) ,b}=(1+K_{b})\nu _{0}+\frac{\left( \nu _{c}-\nu
_{a}\right) ^{2}}{12\nu _{0}\left( 1+K_{b}\right) }
\end{equation}

Following the determination of the quadrupole tensor in $%
Sr_{2.5}Ca_{11.5}Cu_{24}O_{41}$ \cite{Magishi98}, the second term in Eq. (2)
amounts to $42kHz$ ($400$ppm) with a negligible temperature dependence as
compared to the temperature dependence of the first contribution  since ${\left( \nu _{c}-\nu
_{a}\right)}$ remains practically temperature independent \cite
{Takigawa98}.

The pressure dependence of the quadrupolar contribution is not easy to
estimate. However, considering the $100$ ppm change of the quadrupolar
contribution at $B=9.3T (\parallel b)$ which is observed upon $Ca$ substitution between $%
x=0$ and $x=11.5$ in $Sr_{14-x}Ca_{x}Cu_{24}O_{41}$\cite
{Takigawa98,Magishi98} and the expected equivalence between the chemical
pressure of $Ca$ and the applied pressure which is a result of the present study we can
conclude that a change of the quadrupolar term exceeding $100$ ppm in $32 kbar$ is very
unlikely .

The $^{63}Cu$ Knight shift is very anisotropic in spin ladders. Hence, a
precise orientation of the single crystal in the magnetic field along the
corresponding axis is requested for the determination of the actual $%
^{63}K_{\alpha }(T)$ temperature dependence. The situation $B=9.3T
(\parallel b)$ which maximizes the value of the total shift has been reached
through a fine adjustment of the angular position of the pressure cell in
the magnetic field.

\bigskip

Under some circumstances, line shifts and broadenings  at low temperature  both exceed
largely $50 kHz$. 
Since  the $^{63}Cu(metal)$ NMR line does not broaden significantly below $50$ K
under any pressure, the drastic broadening of lines taking place at low $T$ 
cannot be explained by the external field generated by the piston. This
broadening is thus an intrinsic property of ladders. Very similar features have been reported in the NMR
study of $Zn$-doped
$SrCu_{2}O_{3}$
\cite{Fujiwara98}.

\bigskip

 Measurements of the spin-lattice relaxation time $T_{1}$ have been
performed in the same field of $B=9.3T
(\parallel b)$ by the saturation recovery method. The recovery of the
longitudinal magnetization $M_{z}(t)$ towards the equilibrium magnetization $%
M_{\infty }$ was fitted by the expression\cite{Narath67},

\begin{equation}
M_{z}(t)=M_{\infty }-(M_{\infty }-M(0))\left\{ A\exp (-\frac{t}{T_{1}}
)+(1-A)\exp (-\frac{\lambda t}{T_{1}})\right\}
\end{equation}
 Parameters $A$ and $\lambda $ are equal to 0.1 and
6 respectively for the central transition of a spin $I=\frac{3}{2}$ with
non-zero electric field gradients ($EFG$) when the relaxation is purely
magnetic \cite{Narath67}.

\smallskip

The transverse relaxation was measured by the spin-echo decay method and the
spin echo amplitude as a function of the time $\tau $ between the first $\pi
/2$ and the second $\pi $ pulse was fitted above $150K$ by the following
expression, 
\begin{equation}
E(2\tau )=E(0){\rm exp}[-2\tau /T_{2L}-0.5(2\tau /T_{2G})^{2}]
\end{equation}
where $T_{2G}$ is the Gaussian time associated with the nuclear spin-spin
coupling through the conducting electrons and $T_{2L}$ is the Lorentzian
spin-echo time which reads, $1/T_{2L}=3(1/T_{1})_{b}+(1/T_{1})_{a,c}$ .

\section{ Results}

\subsection{\bf \ Knight shifts}

\bigskip



 Figures.1-3 show the $T$
dependences of the $^{63}Cu$ NMR shift $^{63}K_{b}$ in samples $Sr_{14-x}Ca_{x}Cu_{24}O_{41}$ and
$La_{5}Ca_{9}Cu_{24}O_{41}$
 under ambient and also at high pressures after subtraction
of the small quadrupolar contribution in Eq. (2).

The NMR shift consists of two contributions; the
orbital $K_{orb}$ and spin $K_{s}$ (Knight shift) contributions, 
\begin{equation}
K_{b }(T)=K_{b ,orb}+K_{b ,s}(T)
\end{equation}
which can  both be pressure  and temperature dependent although no temperature
dependence is expected for the orbital contribution. The second term
$K_{b ,s}(T)$ is proportional to the uniform spin susceptibility $\chi _{s}(T)$, 
\begin{equation}
K_{b ,s}(T)=A_{b }(q=0)\chi _{s}(T).
\end{equation}
The hyperfine form factor for the $Cu_{2}O_{3}$ ladders, $A_{b }(q)$,
can be written as,

\begin{equation}
A_{b }(q)=A_{b }+2B_{b }\cos q_{x}a+B_{b }\cos q_{y}a
\end{equation}
\begin{equation}
A_{b }(q=0)=A_{b }+3B_{b }
\end{equation}
where $A_{b }$ and $B_{b }$ are the on-site and supertransferred
hyperfine fields, respectively. For a separate estimate of $K_{b,orb}$ and $%
K_{b,s}$ we assume $K_{b,s}=0$ at $T=0$ due to the presence of a spin gap in the
spectrum of spin excitations. Then, $K_{b,orb}=K_{b}(T=0)$.

A previous NMR study of $Sr_{2}Ca_{12}Cu_{24}O_{41}$ under pressure \cite
{Kitaoka99,Mito99} has argued that the  pressure induced increase
of the Knight shift at low temperature with the concomitant suppression of the spin
gap announced in our preliminary works \cite{Mayaffre98,Piskunov00} could actually be
ascribed to the orbital part of the shift decreasing under pressure. It was proposed
that the reduction of the orbital shift could be due to a reduction of the number
of holes in the $Cu-3d_{x_{2}-y_{2}}$ orbital being transferred from $Cu$ to $O$
sites under pressure.

All our data of NMR shifts show a similar line position (between $1.32\%$ and $1.39\%$) at low temperature
under ambient pressure irrespective of the amount of $Ca$ substitution for the entire
$Sr_{14-x}Ca_{x}Cu_{24}O_{41}$ series, (see figures.1a,2a), \cite{Osafune97}. Unlike
the data presented in reference\cite{Mito99} the pressure dependence of the NMR shift is much larger
at low temperature than at room temperature $(0.18\%) {\it  resp} (0.05\%)$ at $36 kbar$, see
figure.3a and therefore cannot be ascribed to a temperature
independent pressure dependence of the orbital shift. In addition, according
to the data presented in reference \cite{Mito99} we should have obtained a
reduction of $0.18\%$ for $K_{b,orb}$ at $P=36kbar$. Our results of figure.3a would thus impose the spin
part
$K_{b,s}$ to be reduced by a factor about two at room temperature and $36kbar$, an assumption which we find
unlikely. We can thus claim that all pressure dependences of NMR shifts for the $%
Sr_{14-x}Ca_{x}Cu_{24}O_{41}$ series are attributable to the spin part.

Interestingly, we have found that $K_{b,orb}$ in $La_{5}Ca_{9}Cu_{24}O_{41}$,fig.1a, 
$(K_{b,orb}=1.49\%)$ is appreciably larger than $K_{b,orb}$ in the $%
Sr_{14-x}Ca_{x}Cu_{24}O_{41}$ series which may be associated with a
difference in symmetry and magnitude of crystalline fields in this compound.

The interesting feature in fig.1a is the
finding of rather similar values  at high temperature and also at low
temperature independent of $x$ and of the applied pressure whereas big changes are observed
in the intermediate temperature regime. For more heavily $Ca$-substituted samples 
$Ca$ (8,9 and 12) , figures.2a,3a the story is quite different since the low
temperature part becomes strongly affected by a change of $x$ and also by the
pressure. We notice an only small pressure dependence of $K_{b}(T)$ at $%
T>150K$ which is {\it at variance} with the strong dependence observed at
lower temperature. Consequently, we can again assume the pressure dependence
of $K_{b,orb}$ to be negligible and the spin contribution to the NMR shift
can be derived by the following equation,

\begin{equation}
K_{b,s}(P,T)=K_{b}(P,T)-K_{b,orb}
\end{equation}

The data of $K_{b,s}$ are shown in figures 1b,2b and 3b. For lightly doped
samples, the pressure (up to $32 kba$r) does not affect the Knight shift at
low temperature (which remains zero). This is {\it at variance} with what is
occurring when the $Ca$ content overcomes $x$=8. Then, the spin part of the
shift becomes pressure dependent at low temperature and a significant upturn of the NMR shift is observed at
very low temperature although essentialy under high pressure. No such effects are observed in lightly
$Ca$-doped samples. The pressure dependence of the zero temperature shift in $Ca12$ fig.3b is indicative for
the appearance of low-lying spin excitations above $20 kbar$. This situation is
reminiscent of the growth of zero frequency spectral weight which is
observed in the undoped two-leg ladders upon substitution of $Cu^{2+}$ by
non-magnetic impurities\cite{Martins96,Fujiwara98}.  A marked
Curie-tail associated with the center of the spectrum is observed at low temperature under pressure in all
samples  with 
$x>5$ . Since this Curie contribution is not detected at ambient
pressure in the same samples we feel confident that it cannot be attributed to
impurities or instrumental shifts of the magnetic field. There is also an
intermediate regime in which $K_{b,s}(T)$ is activated. The scope of the present paper will
be limited to the temperature regime in which  Knight shifts are
activated. An other paper will be devoted to  low lying spin excitations
and the occurence of low temperature line shifts and broadenings.

We shall assume that the temperature dependence of $K_{b,s}$ can be ascribed
to the spin excitations above the spin gap $\Delta _{s}$ as for the undoped
two-leg ladders. Hence, we fit the spin part $K_{b,s}(T)$ using the
following Troyer expression, valid in the low temperature domain \cite{Troyer94}, 
\begin{equation}
K_{b,s}(T)=\frac{C}{\sqrt{T}}\exp \frac{-\Delta _{s}}{T}
\end{equation}
The values of $\Delta _{s}$ thus obtained via fitting the data to eq. (10) for $Ca0,Ca2,Ca5,Ca8$ and $Ca12$ at
ambient and under a pressure of $32 kbar$ are displayed in fig.4.
 The effect of pressure is
large at all $Ca$ contents. The relative reduction of $\Delta _{s}$ under
pressure is steadily increasing with the density of holes ({\it vide infra}%
).We note that the data at ambient pressure agree fairly well with those
published in Ref.\cite{Magishi98}. The spin gap value of 840K which is
obtained for $La5$ is significantly larger than the results of undoped $%
Cu_{2}O_{3}$ ladders which are obtained through the $^{17}O$ Knight shift in $%
La_{6}Ca_{8}Cu_{24}O_{41}$, ($\Delta _{s}=510$K) \cite{Imai98} or by
susceptibility in the $SrCu_{2}O_{3}$ structure \cite{Azuma94}. However, these
results fit rather well with our overall doping dependence of the spin gap
in hole-doped ladders pertaining to the $(Sr/Ca)_{14}Cu_{24}O_{41}$ series
which is presented in the present work.

\bigskip
\subsection{\bf \ Spin lattice and Gaussian relaxation}

Temperature dependences of the spin-lattice relaxation rate $^{63}T_{1}^{-1}$
for $B(9.3T)\parallel b$ axis at various pressures are displayed in figs.5a,b and 6. It should be
noted that the recovery curve of the nuclear magnetization was best fitted 
 above $\sim 70K$ according to Eq. (3) with a single $T_{1}$ component
and with  values for the  $A$    coefficients very close to  values expected 
for magnetic relaxation of the central line in the presence of strong quadrupolar
coupling, namely $A= 0.1$.

\begin{equation}
M_{z}(t)=M_{\infty }-(M_{\infty }-M(0))\left\{ A\exp (-\frac{t}{T_{1}}
)+(1-A)\exp (-\frac{6t}{T_{1}})\right\}
\end{equation}

A look at figs.5a,b and 6 reveals at glance the clear existence of several different
regimes for the temperature dependence of the relaxation rate in agreement
with previous studies of hole-doped spin ladders, \cite
{Magishi98,Mayaffre98,Kitaoka98}. In all samples there is an
activated regime in  ranges depending on the $Ca$ content which
breaks down at low $T$. Below 30K we can clearly see the emergence under 
pressure of a contribution to $1/T_{1}$ which is linear in
temperature (Korringa-like) when $x$ exceeds the value 8. This low temperature domain has been
considered earlier in ref\cite{Mayaffre98} and will be the subject of a further reinvestigation. Moreover,
$1/T_{1}$ 's for all $Sr_{14-x}Ca_{x}Cu_{24}O_{41}$ compounds become nearly $T$%
-independent at high temperature and show a tendency to increase   under
pressure.

\bigskip

Figures 7a,b display the $T$ dependence of the Gaussian component of the
spin-echo decay rate $1/{T_{2G}}$  measured with $B\parallel b$ at different pressures.
We observe that $1/{T_{2G}}$ increases gradually on cooling and is
suppressed by increasing pressure and doping. The determination of $1/{T_{2G}%
}$ is no longer possible below 150K in the heavily doped samples, $(x\geq 5)$%
. This is due to a drastic shortening of $T_{2}$ at low temperature which
prevents the derivation of the Gaussian component.

\section{Interpretation}

\subsection{The isolated spin ladder model}

\subsubsection{Knight shifts} 
The purpose of this section is to examine how the observed Knight shifts 
 can be related to the spin excitations of the $%
(Sr/Ca)_{14}Cu_{24}O_{41}$ spin ladders since $K_{b,s}(T)$ 
is  likely to be governed by these  excitations. There are first the spin excitations 
present in the undoped limit, {\it i.e.} excitations localized on the 
rung triplets in the strong coupling limit  which give rise to  a degenerate
one-magnon band  starting at the energy $\Delta _{s}$ for $k\cong (\pi ,\pi )$ when
the coupling along the legs is switched on. There is also  a two-magnon continuum
starting at the energy
$J$, (see fig.8). These magnon excitations are indeed very robust against the
weakening of the rung coupling since at intermediate coupling $J_{\perp
}/J_{\parallel }\approx 0.5$ the spin gap is still of order $0.2J_{\parallel }$
\cite{Barnes93}. Furthermore the magnon gap is also robust against doping up to 0.25
hole per $Cu$ site {\it albeit} a significant decrease is observed as expected from
the theory \cite{Noack94}. There is also an other kind of triplet spin excitations which must be considered
when spin ladders are doped. They consist of  hole pairs on the same rung dissociating into two individual
quasiparticles on different rungs\cite{Tsunetsugu94}. The role of  quasiparticles on the 
 NMR data will be discussed elsewhere. 

In the gapped regime we shall assume  that low lying excited states correspond to the excitations of single
magnon modes with $k_{x}\approx \pi ,\ k_{y}=\pi $ at the energy $\Delta
_{s} $. These modes govern the temperature dependence of the susceptibility,
leading to the Troyer formula \cite{Troyer94}. We may notice as pointed out in ref\cite{Troyer94} that the
formalism  leading to Eq. (10) applies only to the very low temperature regime.In
particular, the repulsion between magnons which prevent two magnons from occupying the same rung is not
properly taken into account and is essential to get a correct temperature dependence.This may be a reason
explaining why Troyer's formula is followed only up to about $\Delta _{s}/2$ in figs. 1b and 2b.

Values for the spin gap derived from a
fit of the experimental data in figs. 1b,2b,3b with the Troyer formula are displayed  in
figure 4 {\it versus} $Ca$ content at $P=1bar$  and $P=32kbar$
. Figures 5a,b,6 tell us that the spin gap is steadily decreasing upon
$Ca$ doping  as long as $x$ is smaller than 8.
As established by the optical spectral weight data, $Ca$ doping amounts to a
redistribution of the holes between chains and ladders \cite{Osafune97}. 

The main effect of
$Ca$ doping on the structure is a reduction of the $b$ parameter. A similar
conclusion is reached by analyzing the effect of hydrostatic pressure. This
is the $b$ parameter which is most sensitive to pressure. Hence, using the
spin gap {\it versus} $x$ displayed in fig. 4 and structural data under pressure \cite{Pachot99}, we are in
a position to establish a correspondence between the respective roles of pressure and calcium
doping. The contraction of the lattice parameters induced by a pressure of $
P=30kbar$ corresponds to the additional substitution of three $Sr$ by three $Ca$
, {\it i. e.} $b(x,P=30kbar)\approx b(x+3,P=1bar)$ \cite{Pachot99}. We can
attribute the decrease of $\Delta _{s}$ upon Ca substitution in fig. 4 to
 a transfer of  holes from the chains to the ladders.
We can also use the doping calibration curve to plot the variation of $
\Delta _{s}$ against the hole density in the ladders. This plot is displayed
in fig. 9 where  spin gaps at each $Ca$ doping levels have been
normalized by the experimental value of the gap in the $La5$ compound at ambient pressure, 840K.
Furthermore, we have assumed a negligible amount of holes in the ladders of the $La5$ sample under
ambient pressure, given the average $Cu$ valence of 2.04 and the distribution of holes between chains and
ladders in $Ca0$ which supports the dominant role of holes in the chains.
 Subsequently
all gaps have been normalized to the undoped ladder value.

 The amplitude of the
intraladder coupling although unknown in
$La5$ should not depart strongly from the values obtained in layered cuprates or the undoped $
SrCu_{2}O_{3}$ladder, namely $J_{\parallel }\approx 1600K$ leading in turn to $
\Delta_{s}/J_{\parallel}=0.5$. The numerical results for the Heisenberg
ladder\cite{Barnes93} would thus correspond to $J_{\perp }/J_{\parallel
}\approx 1$ which is a reasonable value in the undoped ladder limit. This is consistent  with the 
anisotropy ratio which has been derived from the interpretation of the neutron scattering data
in $La6$ introducing a smaller four-spin exchange interaction \cite{Matsuda00}.  As pressure is increased
we can expect some redistribution of the holes to take place between chains and ladders with a maximum
concentration for the holes on the ladders reachable under pressure corresponding to the uniform situation,
namely 
$n_{L}=0.04$. Using the measured value of the spin gap in $La5$, the plot of fig. 9
suggests that the density of holes in the
$La5$ ladders amounts to 
$n_{L}=0.03$ under $32 kbar$. We have also assumed in fig. 9 the ladder 
hole density reaching the saturation
value of 0.25 in $Ca12$ under pressure.

\bigskip

The data in fig. 9 have been compared to the calculation of the gap for a two-chain
Hubbard model using the density matrix renormalization group \cite{Noack94}.
We notice a good agreement between the theory and the experimental
data at ambient pressure. However, the agreement is no longer satisfying at
$32 kbar$ when $x$ is larger than 8.We are thus forced to look for an other source
of pressure dependence for the spin gap in  $Ca$-rich (strongly hole-doped) samples.

\bigskip

We  notice a persistent  discrepancy between the
determinations of the spin gap by NMR and inelastic neutron spectroscopy (INS). Several
attempts to measure the spin gap in pristine and $Ca$-substituted samples of
the $Sr_{14-x}Ca_{x}Cu_{24}O_{41}$ series by inelastic neutron spectroscopy
have provided a gap of about 32-35 meV in all compounds independent of the
Ca concentration \cite{Eccleston96,Eccleston98,Regnault99,Katano199}. The disagreement is even more
striking in $La5,6$ when spin gaps  are measured by the two techniques \cite{Matsuda00}. Since neutron
data have been obtained at low temperature it could be argued that finite temperature effects could affect
the determination of the spin gap via suceptibility and relaxation time measurements via magnon damping
possibly enhanced by the dissociation of quasiparticle pairs \cite{Fujiyama00}. However the comparison
between neutron and susceptibility data remains a major challenge in the physics of doped spin ladders.
Furthermore, no pressure dependence could be detected by neutron spectroscopy in
$Ca$ rich samples 
\cite{Katano299}. What is badly needed now is a neutron measurement of the spin excitations in the
pressure domain $30-40 kbar$.

\bigskip

What is most remarkable in fig.3b is the existence of a finite value
for the Knight shift at low temperature in $Ca12$ at pressures larger than
$20kbar$. This contribution is also visible in  samples with a smaller $Ca$
content although only in the high pressure limit. A small 
contribution is still observed in $Ca8$ under $32kbar$ but the zero temperature Knight shift of  $Ca5$ would
not depart from  zero even under pressure, see fig. 1b. A finite
zero temperature Knight shift is the signature of low-lying spin
excitations. A look at figures. 3b and 4 suggests that the onset of a zero
temperature Knight shift corresponds to a critical value of the spin gap $
\approx 230$K. The low temperature Knight shift has been the subject of an extensive pressure study
only  in $Ca12$ samples. The data show a steady increase of the low temperature
Knight shift above $20 kbar$ going along with a decrease of the spin gap displayed in fig. 4.

\smallskip
 
We have used the same NMR
resonant circuit to look for superconductivity in $Ca12$ under $36kbar$ without any
magnetic field applied. The onset of the tuning frequency shift occurs at 6.7K, see fig.
10. In spite of a small signal due to an NMR circuit which is not optimized for the
detection of superconductivity we take it as the signature of the superconducting ground
state under $36 kbar$. The onset temperature given by AC methods is known
to be one or two degrees below the transition temperature obtained by the
onset of the resistive transition when the transition is not extremely
narrow. Hence this AC result at $36 kbar$ is in fair agreement with the accepted phase
diagram for superconductivity \cite{Nakanishi98}. Most interesting on fig.11 displaying the pressure
dependence of $\Delta_{s}$ and $T_c$ is the coexistence under
$36 kbar$ of superconductivity at 6.7K with a finite spin gap of 80K and a large zero temperature Knight
shift.

\bigskip

\subsubsection{Spin-lattice relaxation time} 

Our first attempt to understand the spin lattice relaxation data presented in figs. 5a,b and 6 will be to
examine how far one can go into the interpretation using  magnon
modes of the undoped spin ladder corresponding to spin triplet excitations.

The single-magnon modes which  govern the temperature dependence of the
susceptibility are located at an energy equal at least to the gap $\Delta
_{s}$ and consequently the energy conservation law cannot be fulfilled in
the scattering process since the nuclear Larmor frequency $\omega _{0}$ is
much smaller than $\Delta _{s}$. The only relevant relaxation channels are
those which are quasi-elastic and involve a scattering between two excited
magnon states. The channel with both magnons in the region $k_{x}\approx \pi
,\ k_{y}=\pi $ with momentum transfer $q\sim (0,0)$, see fig. 8,leads to the relaxation, 
\begin{equation}
\frac{1}{T_{1}}\sim \mid A_{q=0}\mid ^{2}e^{-\frac{\bigtriangleup _{T_{1}}}{T%
}}
\end{equation}
at low temperature ($T\ll \Delta _{s}$) \cite{Troyer94}, with the assumption
of a quadratic dispersion law for the excitations of undoped Heisenberg
ladders in the case of $J_{\perp }\geq J_{\parallel }$.

\bigskip

It is well known that there exists a discrepancy between derivations of gaps 
$\Delta_{T_{1}}$and $\Delta _{s}$. A similar activation energy for $T_1$ and Knight shift is measured in
the undoped organic Heisenberg ladder $Cu_{2}(C_{5}H_{12}N_{2})_{2}Cl_4$ \cite{Chaboussant97} in agreement
with the theoretical predictions\cite{Troyer94,Sagi96}. However, in all doped ladder systems
$\Delta_{T_{1}}$ extracted from $T_{1}$ is about $1.5$ times larger than the gap $
\Delta_{s}$ obtained from $K_{b,s}$, \cite{Azuma94,Ishida94,Tsuji96,Kumagai97,Magishi98,Mayaffre98}. The
possible reasons for the difference between $\Delta_{T_{1}}$ and $\Delta_{s}$ are
under intense discussion. We intend to propose an interpretation which
allows understanding relaxation data on the basis of a spin gap from Knight
shift experiments only.

\bigskip

As emphasized by Troyer et-al \cite{Troyer94} a contribution to $
^{63}T_{1}^{-1}$ at low $T$ should come from the two-magnon process with
momentum transfer $q\sim (0,0)$ which requires the minimum energy $\Delta
_{s}$. However, Sandvik and Dagotto \cite{Sandvik96} have pointed
out the possibility for three-magnon processes with momentum transfer $
q_{x}\approx \pi ,$ $q_{y}=\pi $ to be equally efficient in the relaxation. These processes consist in the
scattering of an excited state belonging to the two-magnon continuum at the energy $2\Delta _{s}$ near $
k_{x}=0,k_{y}=0$ into a state located on the single magnon dispersion branch
with momentum transfer $q_{x}\approx \pi ,\ q_{y}=\pi $, (see fig. 8). This contribution
can be large because the ladder has strong short-range AF correlations \cite
{Katano299,Ohsugi99,Nagata99}. The numerical studies carried out by Sandvik
et al. \cite{Sandvik96} have concluded to contributions $(\frac{1}{T_{1}}%
)_{q=0}$ and $(\frac{1}{T_{1}})_{q=\pi }$ being of comparable magnitudes at
temperatures such as $T\sim \bigtriangleup _{s}$. Furthermore it was shown
in references \cite{Sandvik96,Ivanov99} that the main contribution to $1/{%
T_{1}}$ in the temperature range $T>\Delta _{s}$ comes from three-magnon
processes with momentum transfers $q_{x}\approx \pi ,\ q_{y}=\pi $.

\bigskip

Recently Ivanov and Lee \cite{Ivanov99} have suggested that $^{63}T_{1}^{-1}$
is determined in the experimentally relevant temperature range by the sum of 
$(\frac{1}{T_{1}})_{q=0}$ and $(\frac{1}{T_{1}})_{q=\pi }$ contributions
with comparable weights. In the low temperature limit, ($T<\Delta _{s}$),
they have found the expression for the contribution to $1/{T_{1}}$ coming
from the processes with $q\cong (\pi ,\pi )$ namely, 
\begin{equation}
(\frac{1}{T_{1}})_{q=\pi }\sim \frac{T}{\bigtriangleup _{s}}e^{-\frac{%
2\bigtriangleup _{s}}{T}}.
\end{equation}
Consequently, we have fitted our $1/{T_{1}}$ data by the form, 
\begin{equation}
\frac{1}{T_{1}}=(\frac{1}{T_{1}})_{q=0}+(\frac{1}{T_{1}})_{q=\pi }\propto
C_{0}\exp (-\frac{\bigtriangleup _{s}}{T})+C_{\pi }(\frac{T}{\Delta _{s}}
)\exp (-\frac{2\bigtriangleup _{s}}{T})
\end{equation}

Eq. (14) represents the analytic form of the low $T$ asymptotics of the two
and three-magnons processes which should apply in the temperature domain $
T/\Delta _{s}<0.5$ in which such an approximation is justified. The solid
lines on figs. 5a,b are a fit according to Eq. (14) using the gap value $
\bigtriangleup _{s}$ derived from the Knight shift data and a ratio $\frac{C_{\pi }}{
C_{0}}=70$. The agreement is fairly good in the gapped regime and Eq. (13) leads to $(\frac{1}{
T_{1}})_{q=\pi }=4.76(\frac{1}{T_{1}})_{q=0}$ at $T/\Delta _{s}=0.5$. The
ratio of the two contributions to the relaxation is practically independent
of the $Ca$ concentration as long as low lying excitations are absent. However this interpretation breaks
down when a finite contribution to the Knight shift arises at low temperature (i.e. in $
Ca$-rich samples under pressure) with the concomitant occurrence of a new
source of relaxation.

\bigskip

A calculation of the relaxation has been performed by Naef and Wang \cite
{Naef00} using the transfer density matrix renormalization group method for
AF couplings relevant in spin ladders. These authors were able to separate
the $q=(0,0)$ and $q=(\pi ,\pi )$ contributions and concluded to the
necessary contribution of the $q=(\pi ,\pi )$ channel for understanding the
cross over observed in $^{63}T_{1}^{-1}$ data of undoped ladders. Using the
experimental data $(\frac{1}{T_{1}})_{q=\pi }=4.76(\frac{1}{T_{1}})_{q=0}$
at $T/\Delta _{s}=0.5$ together with an interpolation of the Naef and Wang
calculation, the ratio $\frac{J_{\perp }}{J_{\parallel }}$ would correspond
to $0.9$. This anisotropy is fairly consistent with our data of spin gap
 in $La_{5}Ca_{9}Cu_{24}O_{41}$ (the  half-filled band
situation), $\Delta_{s}=840K$, using the relation between the spin gap and the anisotropy
calculated by the Heisenberg spin ladder model \cite{Barnes93}. This anisotropy ratio
is not however in agreement with the interpretation of the INS experiments
leading to an anisotropy  of $0.72$ \cite{Katano199} in $Ca11.5$ and 0.55 in $Ca0$ \cite
{Eccleston98}. A similar discrepancy between neutron and NMR techniques has
been noticed in the discussion of the Knight shift data. We may
suggest that the spin gap is severely suppressed by  hole doping going
from $La5$ to $Ca12$ samples while the exchange couplings are much less
affected.

\bigskip

According to the theory, \cite{Naef00} the cross-over between gapped and
paramagnetic regimes of isolated chains should be located around $T/\Delta _{s}=0.6$.
With amplitudes for the gaps obtained from the Knight shift data, the cross-over regime
thus derived is in fair agreement with the temperature dependences of $
1/T_{1}$ displayed in figs.7a,b,8 since the relaxation rate becomes nearly
temperature independent at high temperature as expected above the cross-over
temperature.

It can be noted that our results agree well with those of the $^{17}O$ and $
^{63}Cu$ NMR experiments for undoped $La_{6}Ca_{8}Cu_{24}O_{41}$ and $
Sr_{14-x}Ca_{x}Cu_{24}O_{41}(x=0,12)$ obtained by Fujiyama et al.\cite
{Fujiyama00}. As is well known, magnon processes for the momentum
transfer $q\cong (\pi ,\pi )$ do not contribute to $^{17}T_{1}^{-1}$ because
of the filtering of low frequency spin fluctuations at $q\cong (\pi ,\pi )$
by the hyperfine form factor $^{17}G(q)$ for oxygen sites in the ladders
(see Ref. \cite{Imai98,Fujiyama00}). Using the different $q$-dependences of
the hyperfine form factors for $O$ and $Cu$ sites Fujiyama et al \cite
{Fujiyama00} have shown that the relaxation process at the $Cu$ sites is
dominated by spin excitations near $q=(\pi ,\pi )$ even when the temperature
is lower than the spin gap. Moreover, they have revealed that singlet
correlations along the rungs in the heavily doped compound $Ca12$ become
drastically weaker above a certain temperature and that such a behavior is
most likely caused by the dissociation of bound hole pairs.

\bigskip

\subsection{Transverse coupling}

We have seen in Sec.IV.A.1 that the hole doping being a function of the $Ca$
content and also of the pressure can explain the $x$ dependence of $\Delta_s$ for
all $x$ values at ambient pressure and its pressure dependence provided $x$
remains smaller than 8. Consequently for the large $x$ regime, ($x>8$), we must
look for an other mechanism which is likely  to suppress $\Delta_s$
when the pressure is increased. Recently, Imada and Iino \cite{Imada97} have
studied the effect of the interladder coupling, $J_{L}$, on the critical
magnetic properties of spin ladder systems using the scaling theory together
with quantum Monte Carlo calculations. They have shown that the spin gap is
very sensitive to the ratio $J_{L}/J_{\parallel }$. As discussed above, a
large variation of the intrachain coupling constant, $J_{\parallel }$, under
pressure seems to be unlikely. We can use the model of Imada and Iino \cite
{Imada97} for the case of coupled undoped ladders with an isotropic exchange
and no non magnetic impurities. With the value of $J_{\parallel }$ derived
from inelastic neutron data in $Sr_{2.5}Ca_{11.5}Cu_{24}O_{41}$, $
J_{\parallel }=990\pm 160K$ \cite{Katano199} (the argument would not be
ruined by using $J_{\parallel }=1500K$) and the dependence $\Delta _{s}/
J_{\parallel }$ versus $J_{L}/J_{\parallel }$ obtained in Ref. \cite{Imada97}
we estimate that the reduction of $\Delta _{s}$ from $260K$ at $1bar$ to $
70K $ at $36 kbar$ in $Ca12$ corresponds to the increase $J_{L}$ from $220K$
($J_{L}/J_{\parallel}= 0.20$) to $330$K ($J_{L}/J_{\parallel}=0.30$).

We can notice the agreement between these estimates and the predictions made
in Refs. \cite{Rice93,Gopalan94}for the ratio $J_{L}/J_{\parallel }\sim
0.1\div 0.2$ using a perturbation theory. Consequently, the pressure dependence
of $\Delta _{s}$ in the Fig. 11 could possibly be explained by the increase
of $J_{L}$. Thus, it is likely that the spin gap suppression under 
pressure in the $ Sr_{14-x}Ca_{x}Cu_{24}O_{41}$ series results from a combination of
hole density and interladder coupling increase.

\bigskip

\section{The Gaussian relaxation}

The $1/{T_{2G}}$ Gaussian rate in copper oxides is related to the magnetic
correlation length $\xi $ of the spin fluctuations, \cite
{Pennington89,Pennington91}. Furthermore, the actual relationship between $
\xi $ and $T_{2G}$ is determined by the dimensionality of the materials. It
has been found that $T_{2G}\propto \xi ^{-\frac{1}{2}}$ for $1D$ spin chains 
\cite{Sachdev94,Greven96} and $T_{2G}\propto (T\xi )^{-1}$ for $2D$ systems%
\cite{Imai98}. As to  the relation between $T_{2G}$ and $\xi $ dependence in quasi-$1D$
two-leg ladders, Magishi et-al.\cite{Magishi98} have suggested
for the case of an isotropic intra-ladder coupling, $J_{\parallel }=J_{\perp
}$, the following form,

\begin{equation}
T_{2G}^{2}=\alpha +\beta \xi _{ladder}^{-1}.
\end{equation}
The $T$ dependence of $\xi _{ladder}^{-1}$ in two-leg ladders has been
studied by Greven et-al.\cite{Greven96} who have obtained the relation,

\begin{equation}
\xi_{ladder}^{-1}=\frac{2}{\pi }\Delta _{s}+AT\exp \left( -\frac{\Delta _{s}}{T}%
\right) ,
\end{equation}
in the limit $J_{\perp }\leq J_{\parallel }$ where $\Delta _{s}$ is a spin
gap from the Knight shift data. Using Eqs. (15) and (16) the $T$ dependence
of $T_{2G}^{2}$ can be expressed as,

\begin{equation}
T_{2G}^{2}=A_{0}+BT\exp \left( -\frac{\Delta _{s}}{T}\right) 
\end{equation}
where $A_{0}=T_{2G}^{2}(T=0)\sim \xi _{0}^{-1}\sim \Delta _{s}$ {\it i.e.}
the saturation value of the correlation length due to the finite spin gap.

\smallskip

$T_{2G}^{2}(T)$ data as a function of $Ca$ content and pressure are
displayed in Figs.12a,b. Solid lines are the fits to the data using Eq.
(16) and the values of $\Delta _{s}$ derived from Knight shift measurements.
The overall agreement of our data with those of Ref.(\cite{Magishi98}) at
ambient pressure is rather satisfactory. We can also notice a rise of $
T_{2G}^{2}(T=0)$ with increasing $Ca$ content although this growth is not
linear with respect to the $Ca$ doping.

Magishi et al. \cite{Magishi98} have suggested that the spin correlation length $
\xi $ in hole-doped ladders can be affected by the existence of holes in the
ladders. Hence, $T_{2G}^{2}(T=0)\sim \xi _{eff}^{-1} 
=\xi _{h}^{-1}+\xi _{o}^{-1}$ where $\xi _{h}$ is
a correlation length related to the average distance between holes whereas $
\xi _{o}$ is the intrinsic correlation length of the ladder related to the spin gap. $\xi _{o}^{-1}$
is likely to depend on pressure since it is proportional to the spin gap and 
$\xi _{h}^{-1}$ must depend on the hole concentration {\it i.e.} on both $Ca$
content and pressure since we have found in this work an equivalence between $Ca$
doping and pressure. The data of fig. 12a suggest that for lightly $Ca$-doped
samples $(x\leq 5$) $\xi _{h}^{-1}$ and $\xi _{o}^{-1}$ are of similar order
of magnitude with opposite pressure dependences leading in turn to an
accidental insensitivity of $T_{2G}^{2}(T=0)$ to pressure. However, heavily doped samples are in the
dirty limit, $\xi _{h}^{-1}\gg \xi _{o}^{-1}$, and the increase of $T_{2G}^{2}(T=0)$ under
pressure corresponds to an increase in the density of holes, see for
instance the data for $x=8$ in fig. 12b. Consequently, a change of $\Delta
_{s}$by pressure no longer affects $T_{2G}^{2}(T=0)$ in doped compounds.

However,  the $T$ dependence of $T_{2G}^{2}$ in heavily doped compounds
becomes strongly affected by pressure, see {\it for.ex} data of $Ca8$ and $12
$ on fig. 12b. We see that the whole $T$ dependence of $
T_{2G}^{2}$ for ladders with $x\geq 5$ at high $P$ cannot be fitted by a
single expression such as Eq. (17). The fit of the high $T$ part of the data
(above say $T=250$K) by Eq. (17) leads to the unphysical negative values of $
T_{2G}^{2}(T=0)$ while the fit of data in the $T$ range below $T=250$K gives
reasonable quantities for $T_{2G}^{2}(T=0)$. The increase of $T_{2G}^{2}(T=0)
$ in $Ca5$ and $Ca8$ under pressure is thus mainly governed by  the increase of 
$n_{h}$ in the ladder planes of those compounds. $T_{2G}^{2}(T=0)$ becomes nearly
insensitive to pressure in $Ca12$ as a result of $n_{h}$ reaching saturation
in $Ca12$ under pressure.

\smallskip

Quite recently, Thurber et al.\cite{Thurber00} have shown that the behavior
of $T_{2G}^{2}$ for isolated two- and three-leg ladders at elevated
temperatures is given by,

\begin{equation}
T_{2G}^{2}\propto T^{2}\xi ^{-1}
\end{equation}
Consequently, by combining Eqs.(17) and (18) we obtain the following form
for the description of the high $T$ behavior of $T_{2G}^{2}$,

\begin{equation}
T_{2G}^{2}=A^{}T^{2}+B^{}T^{3}\exp \left( -\frac{\Delta _{s}}{T}\right) 
\end{equation}

As is seen in fig. 12b, $T_{2G}^{2}$ data above $250K$ (at high pressure) can
be rather well fitted by the Eq. (19) (dashed lines). Thus, the $T_{2G}$
data for $x=5,8$ and $12$ compounds under pressure are described by assuming the
existence of a dimensional cross-over at ${T_{cross}}$ for the functional dependence of $T_{2G}^{2}$ with
respect to $\xi _{{}}^{-1}$. The temperature domain below the cross-over
temperature corresponds to a regime of 2-leg ladders where the spin gap is
influenced (decreased) by the inter-ladder interaction. This is relevant only for $Ca8$
and 12 samples under pressure.

\smallskip
The kink which is observed between  solid and dashed lines for $Ca$8 and
12 samples in fig. 12b can be attributed to the manifestation of a cross-over involving the interladder
coupling between the low and high temperature regimes in the $\xi $ dependence of $
T_{2G}^{2}$.

It can be noticed that the values of ${T_{cross}}$ relative to ${\Delta _{s}}$
strongly depends on $x$ \cite{Tcross}. This dependence is understood in
the following manner. $T_{2G}^{2}$ behaves as $T_{2G}^{2}\propto T^{2}\xi
^{-1}$ only above the temperature where the ladders can be considered as  isolated entities, {\it i.
e.} at $T\sim J_{L}$, (the inter-ladder coupling). Therefore, $
T_{cross}$ becomes a function of both $\Delta _{s}$ and $J_{L}$. We can
expect an increase of $J_{L}$ at increasing pressure following the
discussion in Sec. IV.B and also with $Ca$ content leading in turn to an
increase of ${T_{cross}}/{\Delta _{s}}$. The cross-over temperature is such
as ${T_{cross}}$$\geq max(J_{L},\Delta _{s})$. The temperature dependence of 
$T_{2G}^{2}$ in $Ca12$ under pressure, fig. 12b, suggests a cross-over
temperature above 250K which would be in fair agreement with the estimate $
J_{L}\approx 150-300$K derived in Sec. IV.B from the value of the spin gap
taking into account the inter-ladder coupling interaction.

\bigskip

\section {Spin gap and superconductivity}

A  feature of the $P-T$ phase diagram (Figure. 11) that deserves
a special consideration is the existence of a
maximum of  the superconducting  critical temperature when pressure
drives  the characteristic energy scales
for pair binding energy and the spin gap  down to values of the order
of $T_c$. Beyond this optimal
pressure  for the highest $T_c$, denoted as $P_{\rm opt}$, these
intra-ladder gaps  become
irrelevant and hence $T_c$   decreases under  pressure.  The latter pressure
domain is also characterized by a
temperature dependence of resistivity  along the inter-ladder $a$
direction that becomes metallic \cite{Mori97,Auban99}. Such a change of  behavior
in $T_c$ can  be
seen as a cross-over from strong to weak
coupling conditions for  long-range superconducting order.

Arguments in support of this viewpoint follow from  simple
considerations about mechanisms of long-range
superconducting order in coupled ladder systems. Let us consider in
the first place the region below
$P_{\rm opt}$ where the gaps are larger than $T_c$; in this domain,
long-range superconducting coherence
perpendicular to ladder leg direction can only be achieved through a
Josephson coupling. An analysis based on the renormalized perturbation
theory\cite{Bourbonnais91}  shows that the Josephson coupling
takes the form of
\begin{equation}
J_s \approx 2{\xi_s\over c} {t^2_L\over T^*},
\end{equation}
where $\xi_s \approx t_\parallel/T^*$ stands as the size of the
superconducting hole pairs within the
ladders, while
$t_L$ is the inter-ladder single electron hopping ($t_L\ll
t_\parallel$). This expression contains the usual
term ${t^2_L/ T^*}$ for the effective transverse  motion of the pairs
requiring  an intermediate energy of order
$T^*$ (the confinement temperature of hole pairs on  ladders, see Sec.I) and a factor ${\xi_s/ c} $
that  accounts for the increase in probability for pair hopping due to
delocalization  of holes forming the pairs along the ladder
direction\cite{Bourbonnais91}. As for pair
correlations of isolated ladders, they are well known to be strongly
enhanced  in the presence of a spin
gap. Numerical and analytical  results for
$t-J$ and repulsive Hubbard models for example\cite{Hayward95,Schulz96}, show
that  `d-wave' pair susceptibility
follows a power law
$\chi_s(T)\sim (T/\Delta_s)^{-\gamma_s}$ for temperatures below the
spin gap, where $\gamma_s\approx 1$.
A   molecular-field approximation of the Josepshon coupling that
rigourously takes the influence of
intra-ladder correlations into account, allows  one to obtain at once an
expression  for the transition temperature. Thus from the condition
$J_s\chi_s(T_c)=1$ one finds,
\begin{equation}
T_c \sim t^2_L/ T^*,
\end{equation}
which assumes that  $\Delta_s \sim T^*$. It follows that $T_c$  increases
when $T^*$
decreases, a feature that is essentially governed by
the increase of the Josephson coupling under pressure.

Now when pressure is further increased, one reaches the
domain where $T_c\sim T^*$ and a change in
the physical picture of the transition  occurs. The gaps  are no
longer relevant for the
mechanism promoting long-range order and its description   evolves
toward  a weak coupling
picture $-$ a situation that bears close similarities with the
antiferromagnetic ordering
of quasi-one-dimensional organic conductors with a charge gap under
pressure\cite{Bourbonnais99}.  When the
metallic character of the normal state is well developed  along
longitudinal and transverse directions the
motion of the holes  is likely to be coherent along  both directions
leading to  Fermi liquid conditions $-$ or
deconfinement of holes, in agreement with the restoration of a metallic
transverse resistivity.
Thus sufficiently above
$P_{\rm opt}$, the transition temperature
  should follow from a weak coupling BCS prescription which yields
\begin{equation}
T_c \approx t_L e^{-1/|g*|}.
\end{equation}
Here $g^*$ is the residual attractive coupling between deconfined
holes. Since pressure weakens the
strength  of short-range superconducting correlations, it will lead  to  a
decrease of
$|g^*|$  at the energy scale  $t_L$ for
deconfinement\cite{Kishine98,Bourbonnais91}. This  in turn reduces
$T_c$ as a function of pressure. It follows then that  a maximum of
$T_c$  is expected
when the ladder system evolves from  strong to weak coupling
conditions for hole pairing at $P\sim P_{\rm opt}$. A  qualitative description of the pressure profile of
$T_c$ and $\Delta_{s}$ is summarized in Figure. 13.

\bigskip

\section{Conclusion}

The present work has reported a $^{63}Cu$-NMR study of the spin
gap structure and of the spin excitations under pressure of the doped $
Cu_{2}O_{3}$ ladders in the $(Sr/Ca)_{14}Cu_{24}O_{41}$ family covering a
wide range of hole doping. A similar study has been performed with the
compound $La_{5}Ca_{9}Cu_{24}O_{41}$ in which the ladders are nearly
undoped. The spin gap is found to decrease steadily from $La5$ ($\Delta
_{s}=840$K) to $Ca8$ ($\Delta_{s} =300$K) as $x$ is increased and more smoothly
at larger $Ca$ concentration up to $Ca12$. Comparing the effect of pressure
on the spin gap and on the lattice parameters enables us to explain the $x$
and $P$ dependences of the spin gap in the low $Ca$ doping range by the
evolution of the hole doping from 0.028 up to 0.15. A pressure of $32 kbar$
does not alter the spin excitations qualitatively when $x<8$ since the
susceptibility is still reaching a zero value at low temperature. However in 
$Ca$ rich samples, pressure promotes the existence of low lying
spin excitations giving rise to a finite susceptibility at low temperature.

The salient result of this study is the coexistence in the $Ca12$ sample under the pressure of $36 kbar$ of
gapped spin excitations (with a gap of 80K) with superconductivity at 6.7K.
The results of the present work provide a further confirmation to the early study published by Mayaffre
et-al \cite{Mayaffre98} although the actual title of ref \cite{Mayaffre98} may appear as somewhat
misleading in the light of coexisting gapped and low lying spin excitations.

These new results are supporting the exciting
 prospect of superconductivity induced by the interladder
tunnelling of preformed pairs \cite{Rice93,Kishine97,Kishine98} at least in
the pressure domain where the pressure remains smaller than the pressure
corresponding to the maximum of the critical temperature for
superconductivity.

After the study of the spin-lattice relaxation of $^{63}Cu$ nuclei in the $
Sr_{14-x}Ca_{x}Cu_{24}O_{41}$ series for various levels of $Ca$ substitution
and also under pressure we are able to identify two different temperature
regimes. A fairly consistent interpretation of the spin lattice relaxation
data above 50K is obtained by taking into consideration two and three-magnon
modes relaxation channels responsible for the activated nuclear relaxation
in the gapped temperature regime and a quasi-1D spin dynamics at high
temperature. Using a linear combination between two and three magnon modes
enables us to remove the discrepancy between activation energies obtained
from susceptibility and $1/T_1$ experiments.

The  dependence of the Gaussian relaxation time has
confirmed that  the rise of the correlation length at low
temperature is limited by two factors; the quantum limitation due to the finite spin gap and  the average
distance between holes. It is the second factor which prevails in $Ca$-rich samples under pressure. A kink in
the temperature dependence of
$1/T_{2G}$ at high pressure in  heavily doped samples has been attributed to the manifestation of a cross-over between the behavior of isolated ladders at high temperature and the
behavior of coupled ladders  at low
temperature.

Doped cuprate spin ladders in which the doping can be controlled by pressure
are remarkable prototype compounds for the study of the mechanism for
superconductivity in cuprates. The interplay between the real space pair
binding and superconductivity can also be studied. Experiments at even
higher pressures where a more conventional kind of superconductivity is
expected would be most useful.

\section{Acknowledgments}

Y.P acknowledges the French Ministry of Research and CNRS for a postdoctoral financial support. We
thank J.P.Boucher for a useful discussion.  A BQR grant from the Universit\'e Paris-Sud  is also
acknowledged.

\newpage
\begin{center}
FIGURE\ CAPTIONS
\end{center}
\bigskip

{\bf Figure 1:} a) Temperature dependences of the $^{63}Cu$ NMR shift $
^{63}K_{b}$  in  $La_{5}Ca_{9}Cu_{24}O_{41}$ ($La5$) and $
Sr_{14-x}Ca_{x}Cu_{24}O_{41}$ ($x=0,2,5$) for $H\parallel {\bf b}$ under
ambient and high pressures. b) $T$ dependences of the $^{63}Cu$ Knight shift $^{63}K_{b,s}$
in $La_{5}Ca_{9}Cu_{24}O_{41}$ ($La5$) and $Sr_{14-x}Ca_{x}Cu_{24}O_{41}$ ($
x=0,2,5$) for $H\parallel {\bf b}$ under ambient and high pressures. The
lines are the fits to the data using Eq. (10).

\bigskip
{\bf Figure 2:} a) Temperature dependences of the $^{63}Cu$ NMR shift $
^{63}K_{b}$ in   $
Sr_{14-x}Ca_{x}Cu_{24}O_{41}$ ($x=8,9$) for $H\parallel {\bf b}$ under
ambient and high pressures. b) $T$ dependences of the $^{63}Cu$ Knight shift $^{63}K_{b,s}$
in $Sr_{14-x}Ca_{x}Cu_{24}O_{41}$ ($x=8,9$) for $H\parallel {\bf b}$ under
ambient and high pressures. The lines are the fits to the data using Eq. (10).

\bigskip
{\bf Figure 3: }a) Temperature dependences of the $^{63}Cu$ NMR shift $
^{63}K_{b}$ in $Sr_{2}Ca_{12}Cu_{24}O_{41}$ for $H\parallel {\bf b}$ at
different pressures. b) $T$ dependences of the $^{63}Cu$ Knight shift $^{63}K_{b,s}$
in $Sr_{2}Ca_{12}Cu_{24}O_{41}$ for $H\parallel {\bf b}$ at different
pressures. The lines are the fits to the data using Eq. (10).

\bigskip
{\bf Figure 4:} Dependence of the spin gap $\Delta _{s}$ upon $Ca$ substitution
under ambient and high pressure. The values of $\Delta _{s}$ were obtained
from the fits to the Knight shift data using Eq. (10). The solid lines are a
guide for the eye.

\bigskip
{\bf Figure 5:} a) Temperature dependences of the spin lattice relaxation rate $%
^{63}T_{1}^{-1}$ in $La_{5}Ca_{9}Cu_{24}O_{41}$ ($La5$) and $%
Sr_{14-x}Ca_{x}Cu_{24}O_{41}$ ($x=0,2$) for $H\parallel {\bf b}$ under
ambient and high pressures. The lines are the fits to the data using Eq. (6). b) Temperature dependences
of the spin lattice relaxation rate $^{63}T_{1}^{-1}$ in $Sr_{14-x}Ca_{x}Cu_{24}O_{41}$ ($x=5,8,9$) for
$ H\parallel {\bf b}$ under ambient and high pressures. The lines are the fits
to the data using Eq. (14).

\bigskip
{\bf Figure 6:} Temperature dependences of the spin lattice relaxation rate $%
^{63}T_{1}^{-1}$in $Sr_{2}Ca_{12}Cu_{24}O_{41}$ for $H\parallel {\bf b}$ at
different pressures. The lines are the fits to the data using Eq. (14). In
the inset, $(T_{1}T)^{-1}$ $vs$ $T$.

\bigskip
{\bf Figure 7:} a) $T$ dependences of the Gaussian component of the spin-echo
decay rate $T_{2G}^{-1}$ in lightly doped $Sr_{14-x}Ca_{x}Cu_{24}O_{41}$ ($%
x=0,2,5$) for $H\parallel {\bf b}$ under ambient and high pressures. b) $T$ dependences of the Gaussian
component of the spin-echo decay rate $T_{2G}^{-1}$ in lightly doped $Sr_{14-x}Ca_{x}Cu_{24}O_{41}$ ($
x=8,12$) for $H\parallel {\bf b}$ under ambient and high pressure.

\bigskip
{\bf Figure 8:} Qualitative picture of the dispersion for $J_{\perp }\gg
J_{\parallel }$ (dashed line) and for $J_{\parallel }=J_{\perp }$ (solid
line). Hatched region corresponds to the continuum of two-magnon states at $%
k_{y}=0$ and with minimum energy $2\Delta _{s}$. Three-magnon processes with
momentum transfer $q_{x}\approx \pi ,\ q_{y}=\pi $ consist in the scattering
of an excited state belonging to the two-magnon continuum at the energy $%
2\Delta _{s}$ near $k_{x}=0,k_{y}=0$ into a state located on the single
magnon dispersion branch (dashed line on the figure).

\bigskip
{\bf Figure 9:} The spin gaps $\Delta _{s}^{\exp }$ and $\Delta
_{s}^{theor} $ normalized to the undoped ladder value $\Delta _{s}(n_{h}=0)$ $La5$
(see text for details) as a function of the hole density $n_{h}$. The values of $
\Delta _{s}^{\exp }$ were obtained for various doping level from Knight
shift measurements under ambient and high pressures. $\Delta _{s}^{theor}$
is the gap calculated for a two-chain Hubbard model using the density matrix
renormalization group by Noack et al\cite{Noack94}. The solid and dashed
lines are a guide for the eye.

\bigskip
{\bf Figure 10:} $T$ dependence of the tuning frequency shift $\Delta f(T)$ of the
NMR resonant circuit caused by a superconducting transition in $
Sr_{2}Ca_{12}Cu_{24}O_{41}$ at $36kbar$. $\Delta f(T)$ has been normalized
to the maximum value of the shift $\Delta f_{\max }=\Delta f(T=4K)$.

\bigskip
{\bf Figure 11:} Pressure dependence of the spin gap $\Delta _{s}$ obtained
from the fits to the Knight shift data using Eq. (10) and of the superconducting transition temperature,
$T_{c}$, for $ Sr_{2.5}Ca_{11.5}Cu_{24}O_{41}$ ($\triangle $)\cite{Nagata97} and for $
Sr_{2}Ca_{12}Cu_{24}O_{41}$ (*)\cite{Mayaffre98,Auban99}.

\bigskip
{\bf Figure 12:} a) $T_{2G}^{2}$ as a function of $T$ in $%
Sr_{14-x}Ca_{x}Cu_{24}O_{41}$ ($x=0,2,5$) for $H\parallel {\bf b}$ at $%
P=1bar $ and $P=32 kbar$. The lines are the fits to the data using Eq. (17)
with $\Delta _{s}$ derived from Knight shift measurements. b) $T_{2G}^{2}$ as a function of $T$ in $%
Sr_{14-x}Ca_{x}Cu_{24}O_{41}$ ($x=8,12$) for $H\parallel {\bf b}$ under
ambient and high pressures. The solid lines are the fits to the data using
Eq. (17). The dashed lines are the fits to the data using Eq. (19).

\bigskip
{\bf Figure 13:} Qualitative phase diagram of the pair formation temperature $
T^{*}$ and superconducting transition temperature $T_{c}$ as a function of the
pressure.


\begin{thebibliography}{10}
\bibitem{Udo}
The compound $La_{6}Ca_{8}Cu_{24}O_{41}$ would have been an even better choice
  to provide the reference undoped ladder system but the sample quality was not
  as good as for the $La_{5}Ca_{9}Cu_{24}O_{41}$ single crystals.
\end{thebibliography}

\newpage

\begin{figure}[ htb]
\epsffile{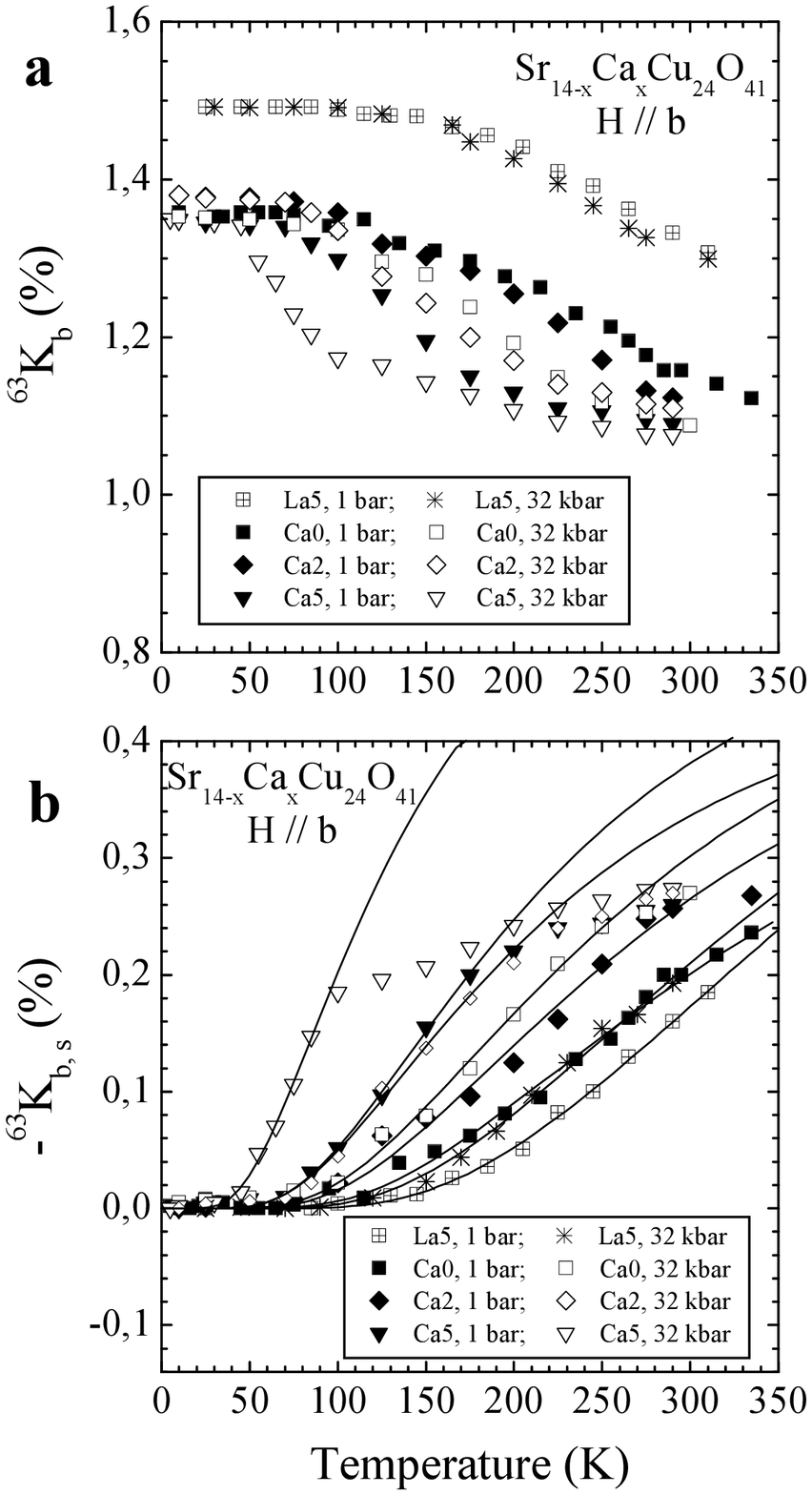}
\caption{ a) Temperature dependences of the $^{63}Cu$ NMR shift $
^{63}K_{b}$  in  $La_{5}Ca_{9}Cu_{24}O_{41}$ ($La5$) and $
Sr_{14-x}Ca_{x}Cu_{24}O_{41}$ ($x=0,2,5$) for $H\parallel {\bf b}$ under
ambient and high pressures. b) $T$ dependences of the $^{63}Cu$ Knight shift $^{63}K_{b,s}$
in $La_{5}Ca_{9}Cu_{24}O_{41}$ ($La5$) and $Sr_{14-x}Ca_{x}Cu_{24}O_{41}$ ($
x=0,2,5$) for $H\parallel {\bf b}$ under ambient and high pressures. The
lines are the fits to the data using Eq. (10).}
\label{}
\end{figure}

\begin{figure}[ htb]
\epsffile{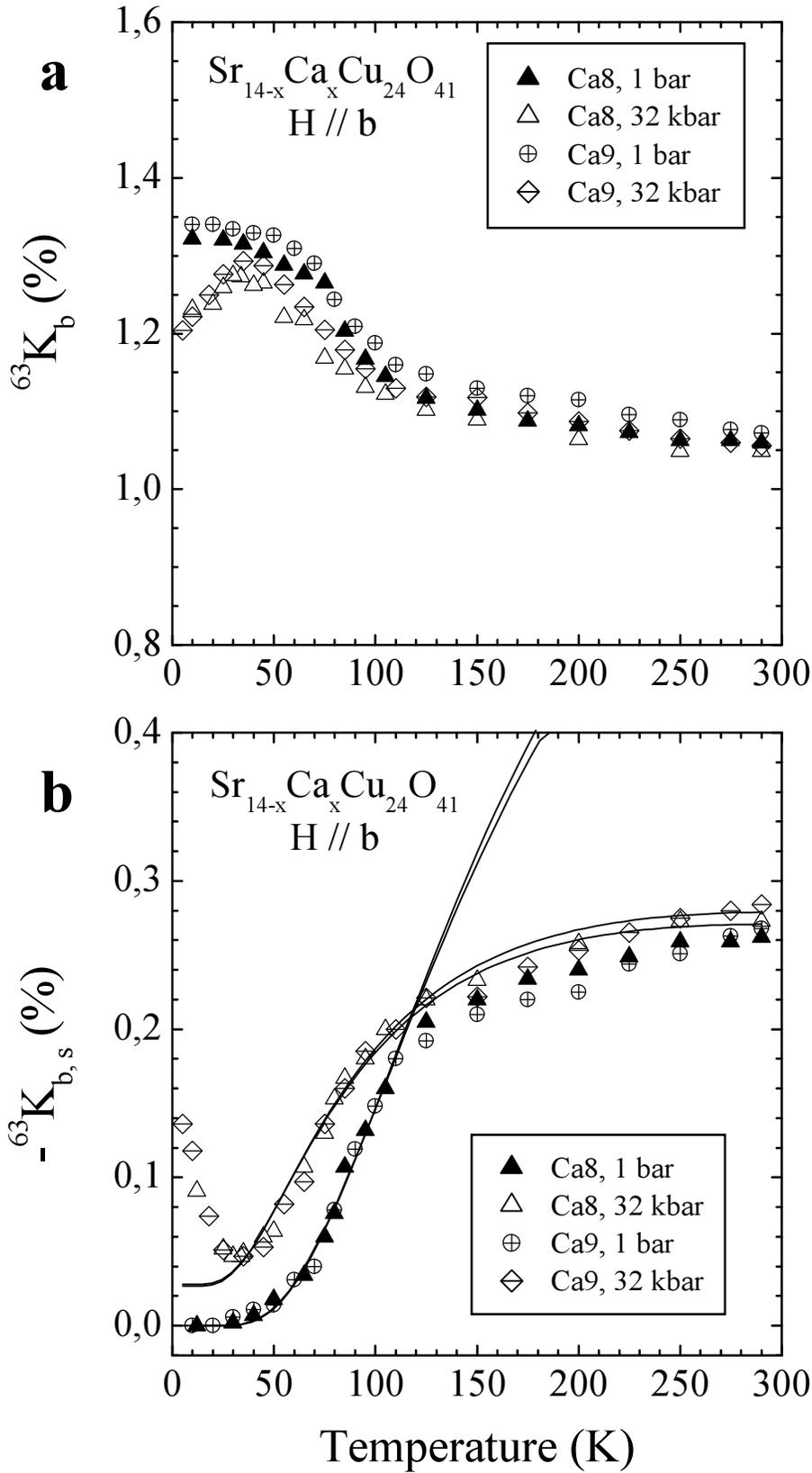}
\caption{ a) Temperature dependences of the $^{63}Cu$ NMR shift $
^{63}K_{b}$ in   $
Sr_{14-x}Ca_{x}Cu_{24}O_{41}$ ($x=8,9$) for $H\parallel {\bf b}$ under
ambient and high pressures. b) $T$ dependences of the $^{63}Cu$ Knight shift $^{63}K_{b,s}$
in $Sr_{14-x}Ca_{x}Cu_{24}O_{41}$ ($x=8,9$) for $H\parallel {\bf b}$ under
ambient and high pressures. The lines are the fits to the data using Eq. (10).}
\label{}
\end{figure}

\begin{figure}[ htb]
\epsffile{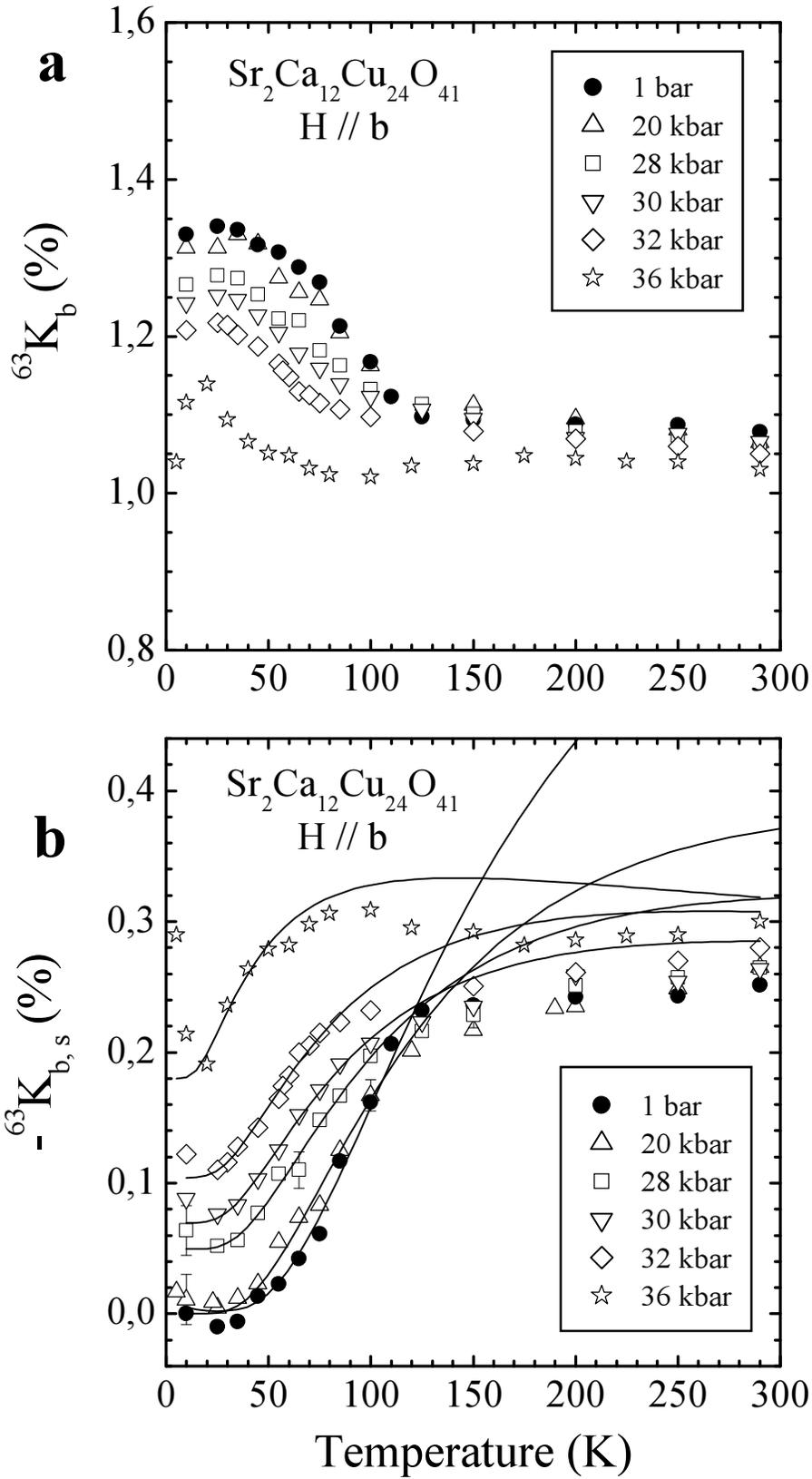}
\caption{a) Temperature dependences of the $^{63}Cu$ NMR shift $
^{63}K_{b}$ in $Sr_{2}Ca_{12}Cu_{24}O_{41}$ for $H\parallel {\bf b}$ at
different pressures. b) $T$ dependences of the $^{63}Cu$ Knight shift $^{63}K_{b,s}$
in $Sr_{2}Ca_{12}Cu_{24}O_{41}$ for $H\parallel {\bf b}$ at different
pressures. The lines are the fits to the data using Eq. (10).
}
\label{}
\end{figure}

\begin{figure}[ htb]
\epsffile{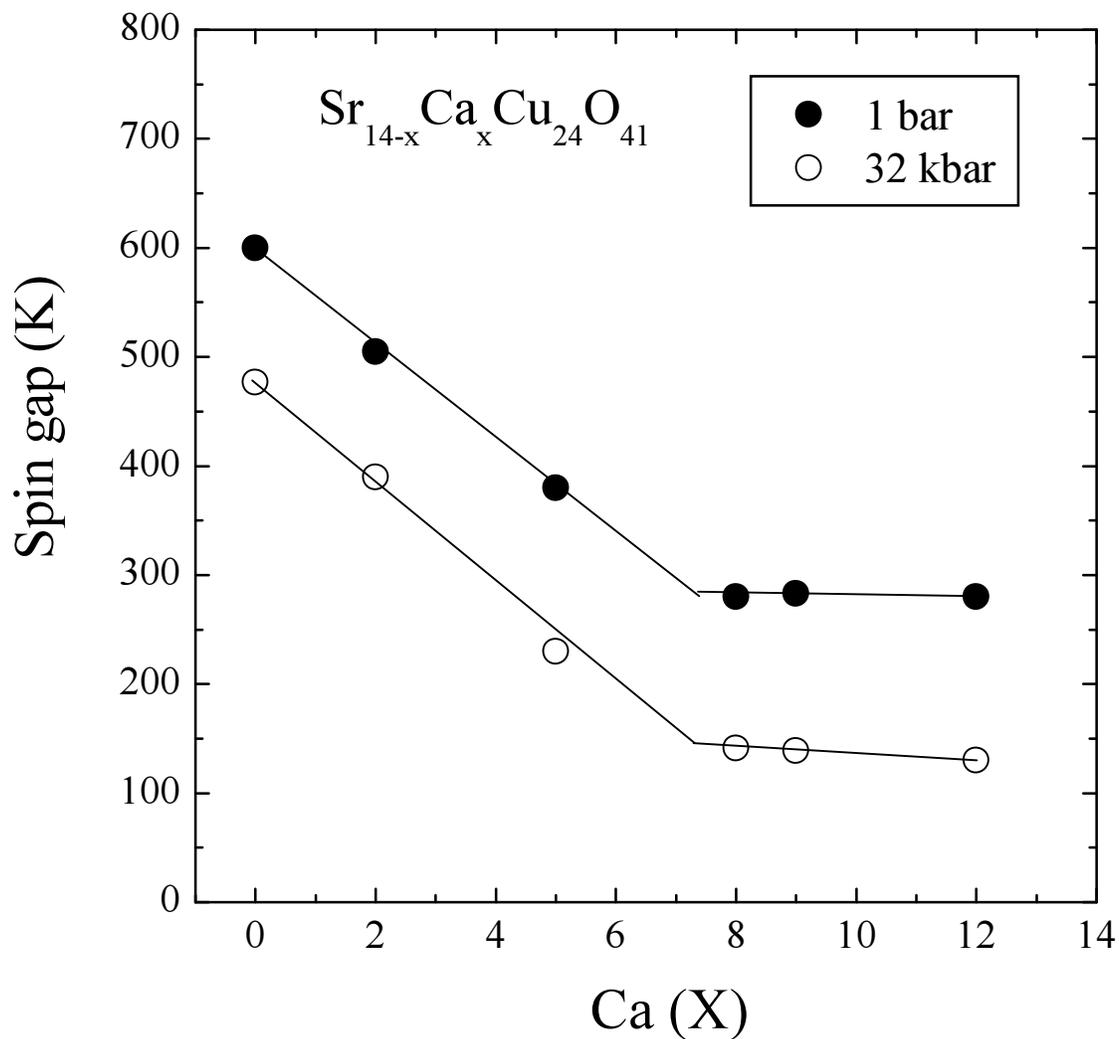}
\caption{ Dependence of the spin gap $\Delta _{s}$ upon $Ca$ substitution
under ambient and high pressure. The values of $\Delta _{s}$ were obtained
from the fits to the Knight shift data using Eq. (10). The solid lines are a
guide for the eye.
}
\label{}
\end{figure}

\begin{figure}[ htb]
\epsffile{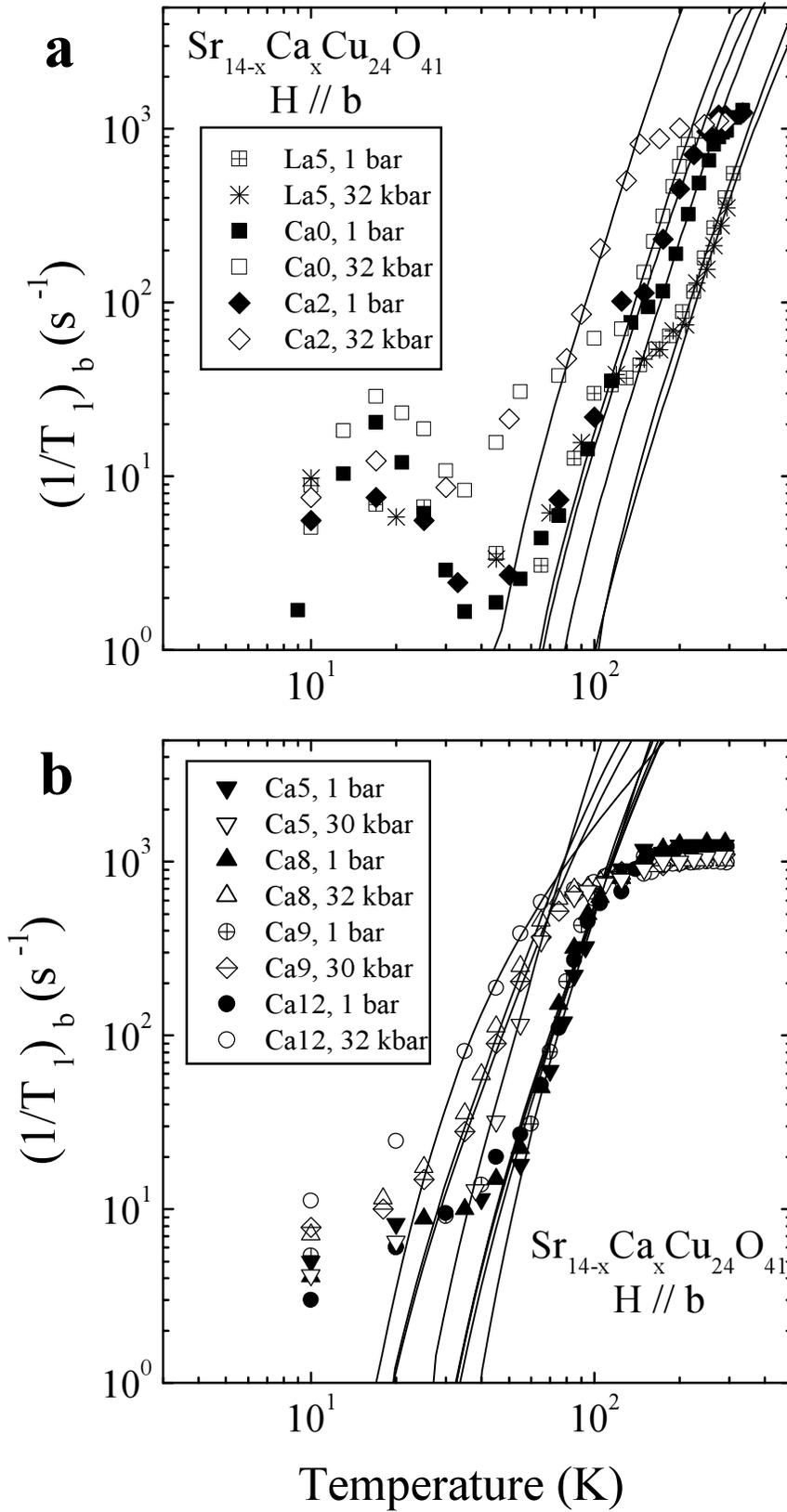}
\caption{a) Temperature dependences of the spin lattice relaxation rate $%
^{63}T_{1}^{-1}$ in $La_{5}Ca_{9}Cu_{24}O_{41}$ ($La5$) and $%
Sr_{14-x}Ca_{x}Cu_{24}O_{41}$ ($x=0,2$) for $H\parallel {\bf b}$ under
ambient and high pressures. The lines are the fits to the data using Eq. (6). b) Temperature dependences
of the spin lattice relaxation rate $^{63}T_{1}^{-1}$ in $Sr_{14-x}Ca_{x}Cu_{24}O_{41}$ ($x=5,8,9$) for
$ H\parallel {\bf b}$ under ambient and high pressures. The lines are the fits
to the data using Eq. (14).
}
\label{}
\end{figure}

\begin{figure}[ htb]
\epsffile{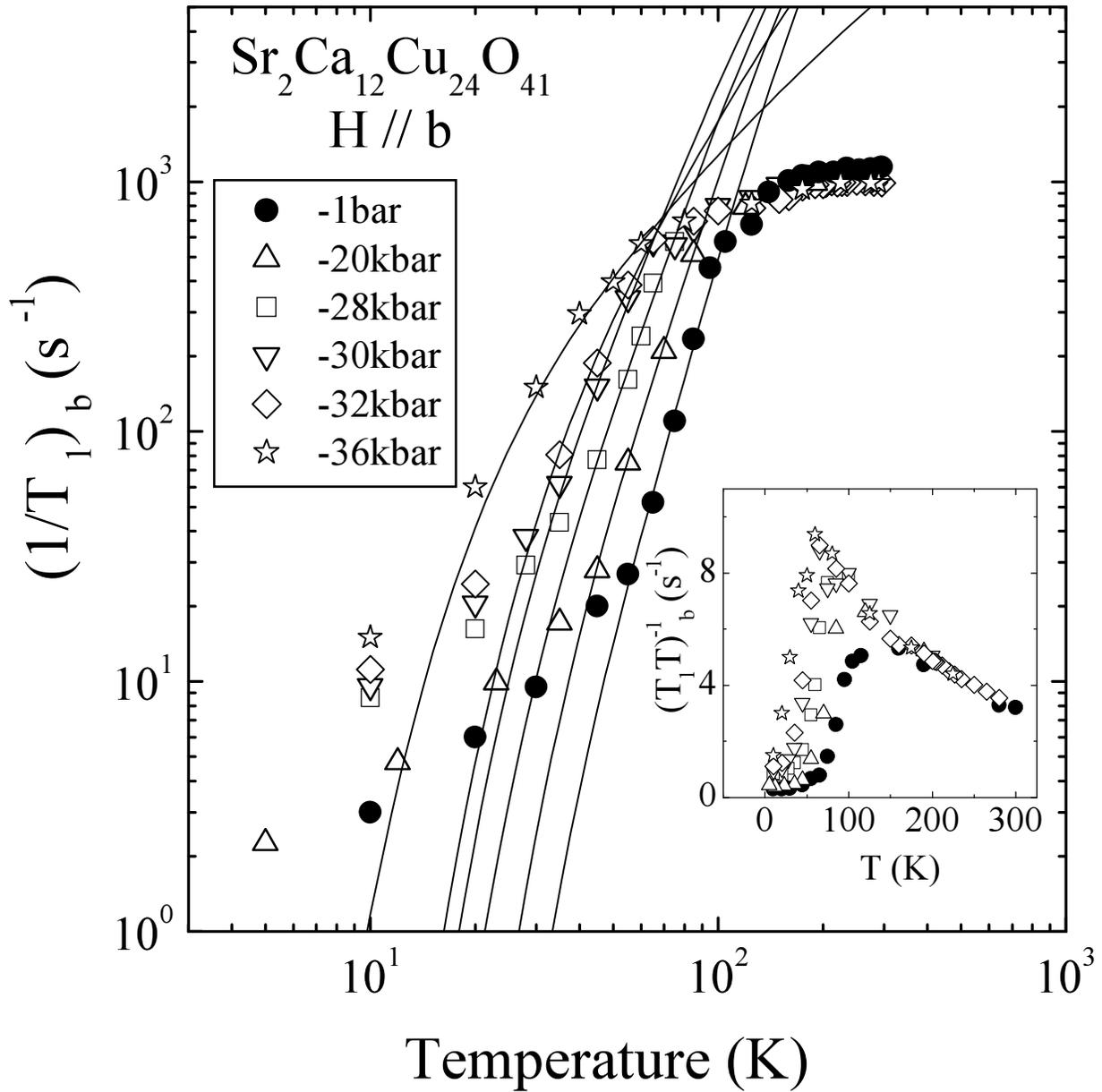}
\caption{ Temperature dependences of the spin lattice relaxation rate $%
^{63}T_{1}^{-1}$in $Sr_{2}Ca_{12}Cu_{24}O_{41}$ for $H\parallel {\bf b}$ at
different pressures. The lines are the fits to the data using Eq. (14). In
the inset, $(T_{1}T)^{-1}$ $vs$ $T$.}
\label{}
\end{figure}

\begin{figure}[ htb]
\epsffile{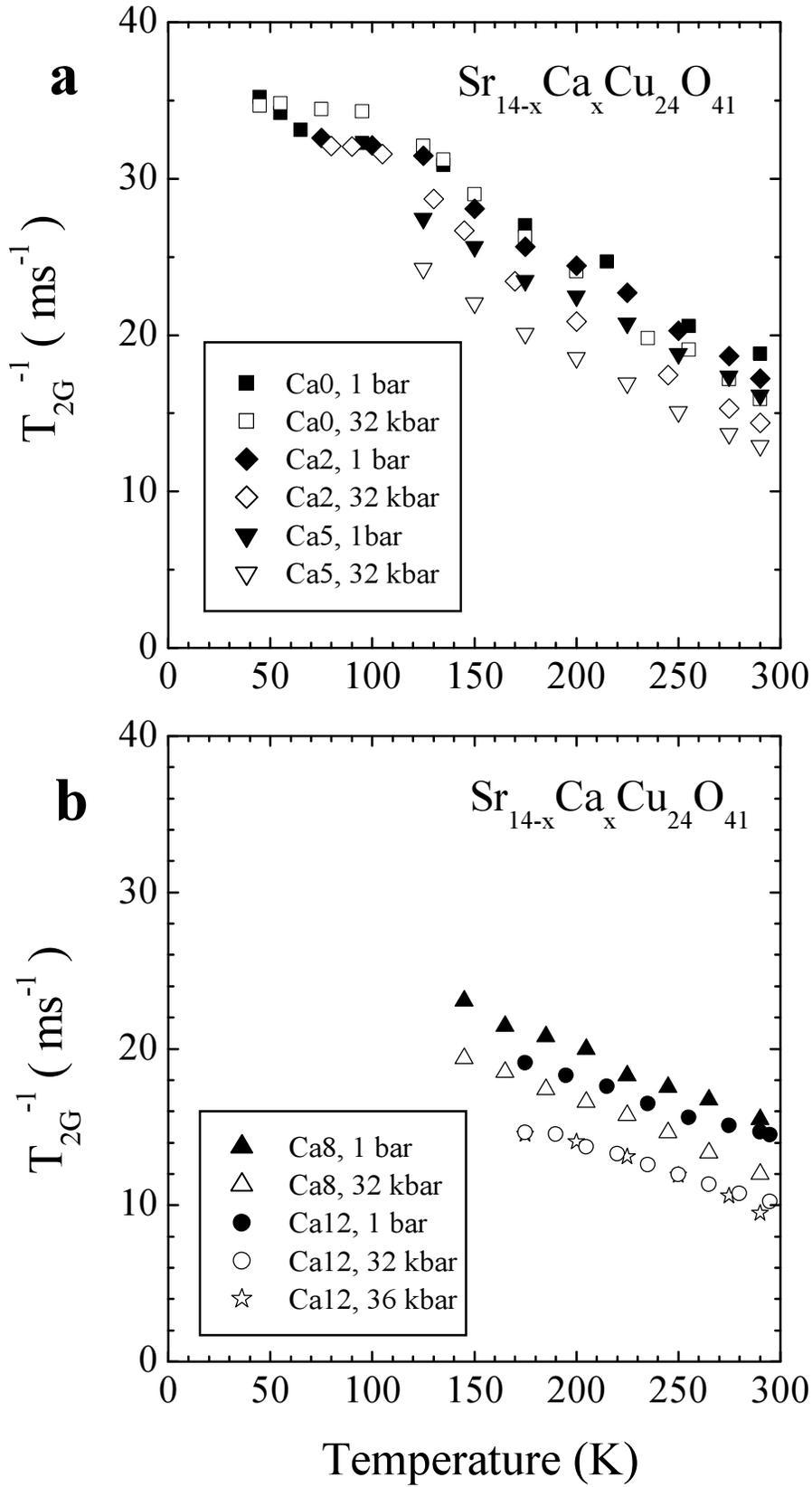}
\caption{ a) $T$ dependences of the Gaussian component of the spin-echo
decay rate $T_{2G}^{-1}$ in lightly doped $Sr_{14-x}Ca_{x}Cu_{24}O_{41}$ ($%
x=0,2,5$) for $H\parallel {\bf b}$ under ambient and high pressures. b) $T$ dependences of the Gaussian
component of the spin-echo decay rate $T_{2G}^{-1}$ in lightly doped $Sr_{14-x}Ca_{x}Cu_{24}O_{41}$ ($
x=8,12$) for $H\parallel {\bf b}$ under ambient and high pressure.}
\label{}
\end{figure}

\begin{figure}[ htb]
\epsffile{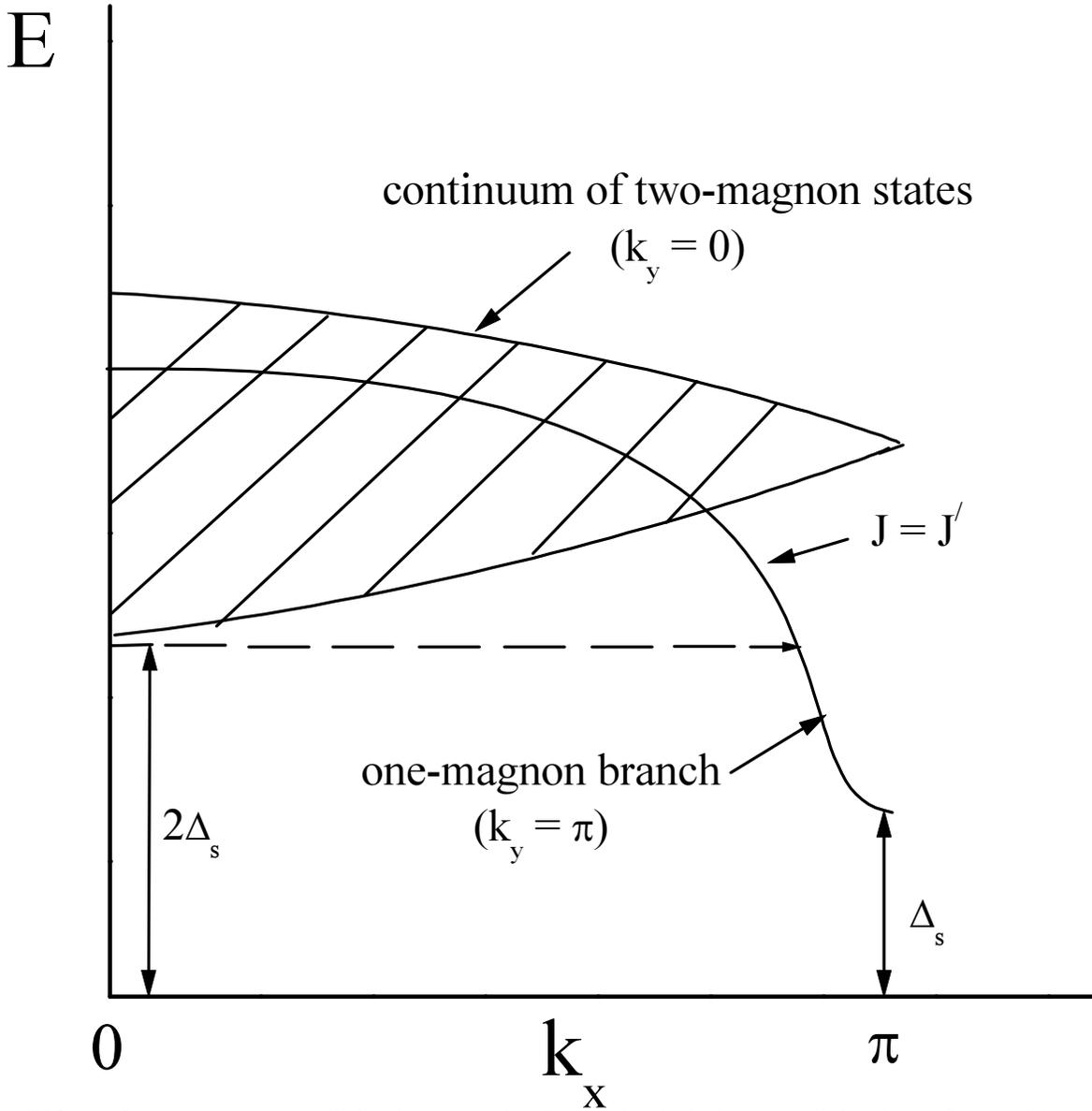}
\caption{Qualitative picture of the dispersion for $J_{\perp }\gg
J_{\parallel }$ (dashed line) and for $J_{\parallel }=J_{\perp }$ (solid
line). Hatched region corresponds to the continuum of two-magnon states at $%
k_{y}=0$ and with minimum energy $2\Delta _{s}$. Three-magnon processes with
momentum transfer $q_{x}\approx \pi ,\ q_{y}=\pi $ consist in the scattering
of an excited state belonging to the two-magnon continuum at the energy $%
2\Delta _{s}$ near $k_{x}=0,k_{y}=0$ into a state located on the single
magnon dispersion branch (dashed line on the figure).
}
\label{}
\end{figure}

\begin{figure}[ htb]
\epsffile{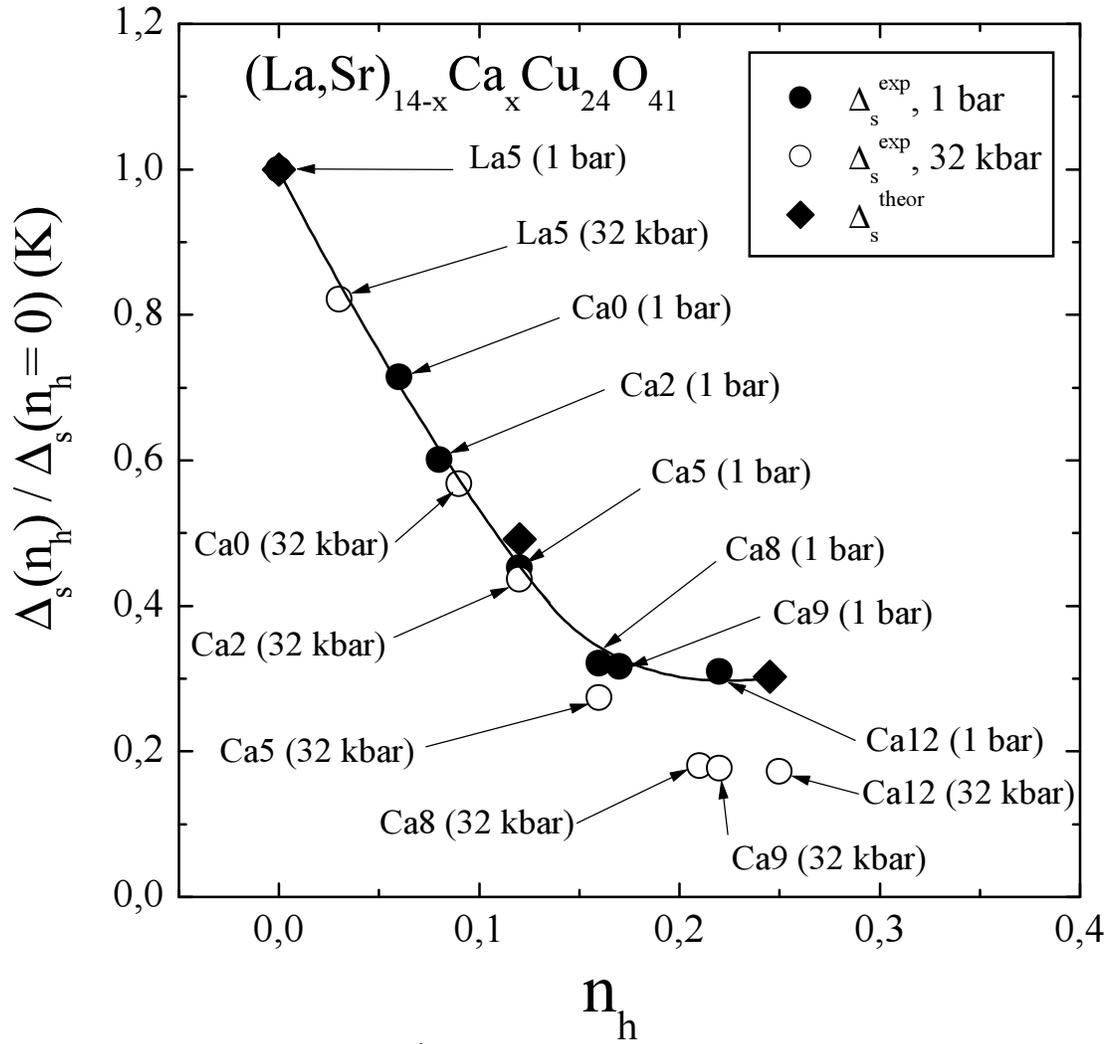}
\caption{The spin gaps $\Delta _{s}^{\exp }$ and $\Delta
_{s}^{theor} $ normalized to the undoped ladder value $\Delta _{s}(n_{h}=0)$ for $La5$
(see text for details) as a function of the hole density $n_{h}$. The values of $
\Delta _{s}^{\exp }$ were obtained for various doping level from Knight
shift measurements under ambient and high pressures. $\Delta _{s}^{theor}$
is the gap calculated for a two-chain Hubbard model using the density matrix
renormalization group by Noack et al \protect\cite{Noack94}. The solid and dashed
lines are a guide for the eye.
}
\label{}
\end{figure}

\begin{figure}[ htb]
\epsffile{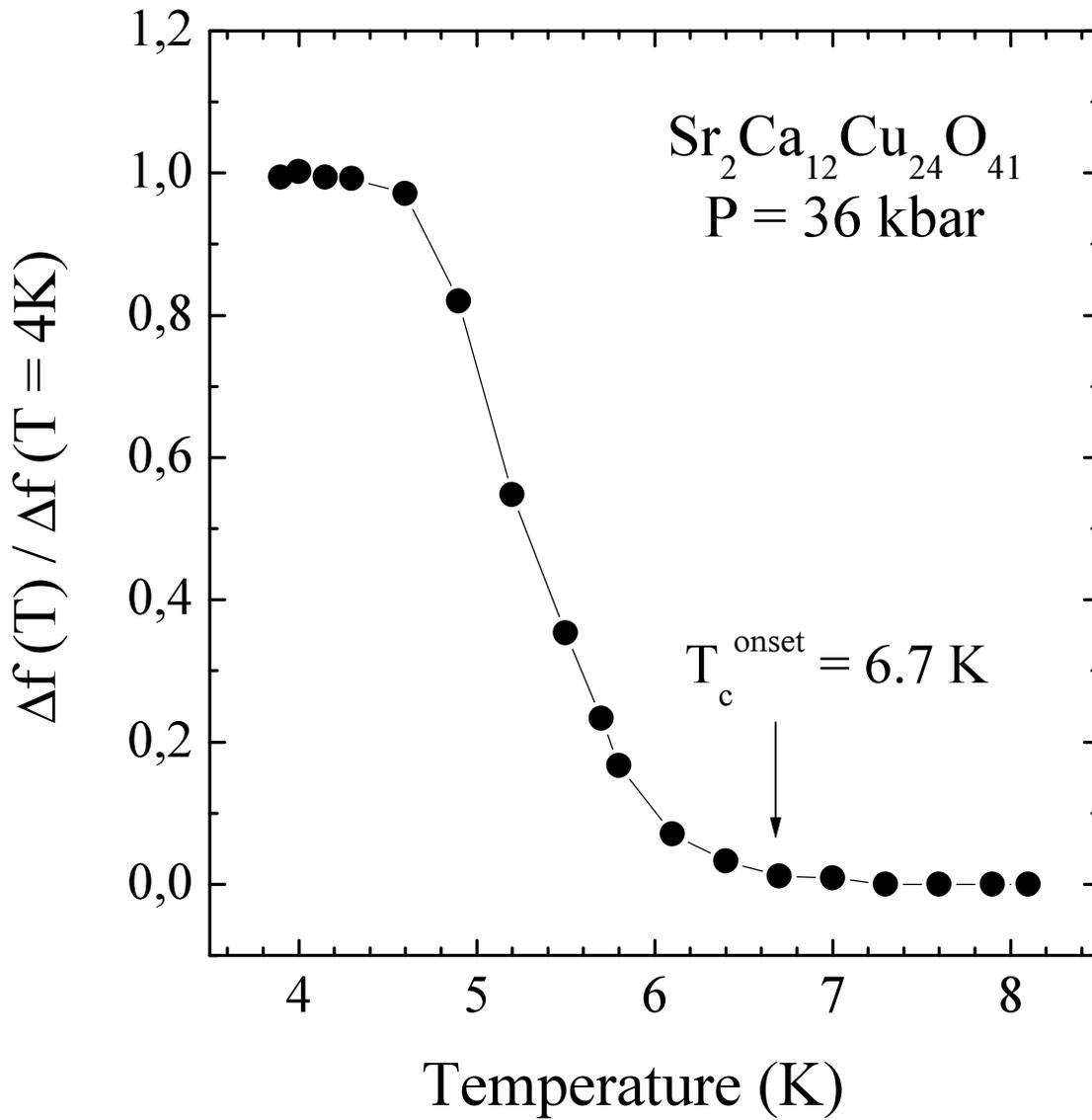}
\caption{ $T$ dependence of the tuning frequency shift $\Delta f(T)$ of the
NMR resonant circuit caused by a superconducting transition in $
Sr_{2}Ca_{12}Cu_{24}O_{41}$ at $36kbar$. $\Delta f(T)$ has been normalized
to the maximum value of the shift $\Delta f_{\max }=\Delta f(T=4K)$.
}
\label{}
\end{figure}

\begin{figure}[ htb]
\epsffile{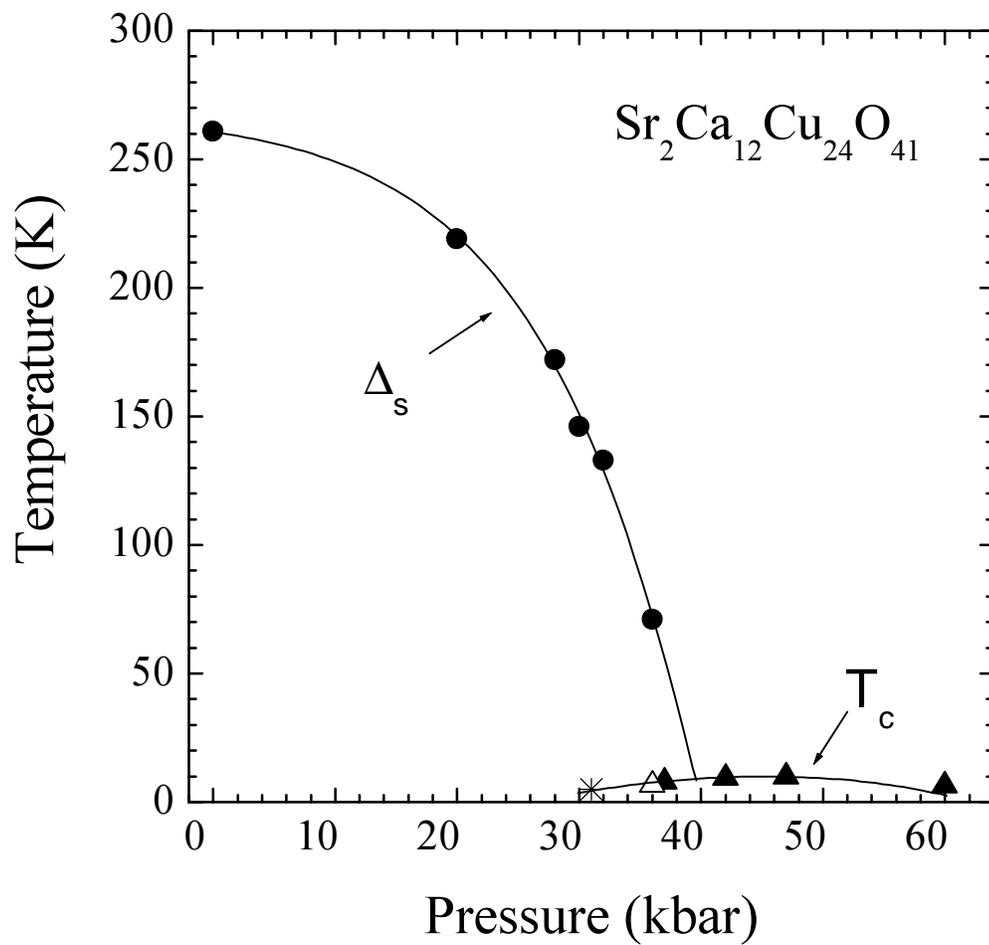}
\caption{Pressure dependence of the spin gap $\Delta _{s}$ obtained
from the fits to the Knight shift data using Eq. (10) and of the superconducting transition temperature,
$T_{c}$, for $ Sr_{2.5}Ca_{11.5}Cu_{24}O_{41}$ ($\triangle $) \protect\cite{Nagata97} and for
$Sr_{2}Ca_{12}Cu_{24}O_{41}$ (*) \protect\cite{Mayaffre98,Auban99}.}
\label{}
\end{figure}

\begin{figure}[ htb]
\epsffile{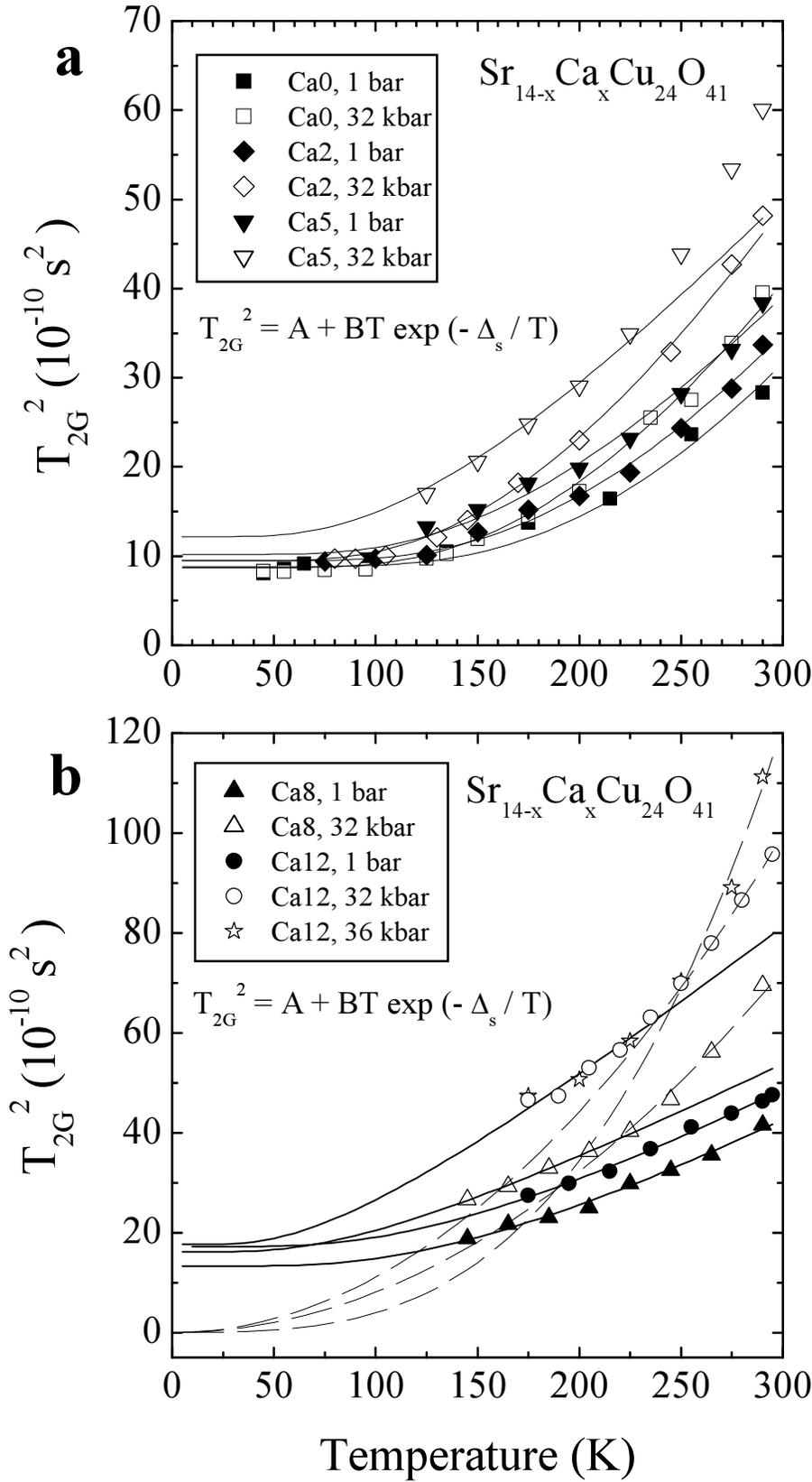}
\caption{a) $T_{2G}^{2}$ as a function of $T$ in $%
Sr_{14-x}Ca_{x}Cu_{24}O_{41}$ ($x=0,2,5$) for $H\parallel {\bf b}$ at $%
P=1bar $ and $P=32 kbar$. The lines are the fits to the data using Eq. (17)
with $\Delta _{s}$ derived from Knight shift measurements. b) $T_{2G}^{2}$ as a function of $T$ in $%
Sr_{14-x}Ca_{x}Cu_{24}O_{41}$ ($x=8,12$) for $H\parallel {\bf b}$ under
ambient and high pressures. The solid lines are the fits to the data using
Eq. (17). The dashed lines are the fits to the data using Eq. (19).
}
\label{}
\end{figure}

\begin{figure}[ htb]
\epsffile{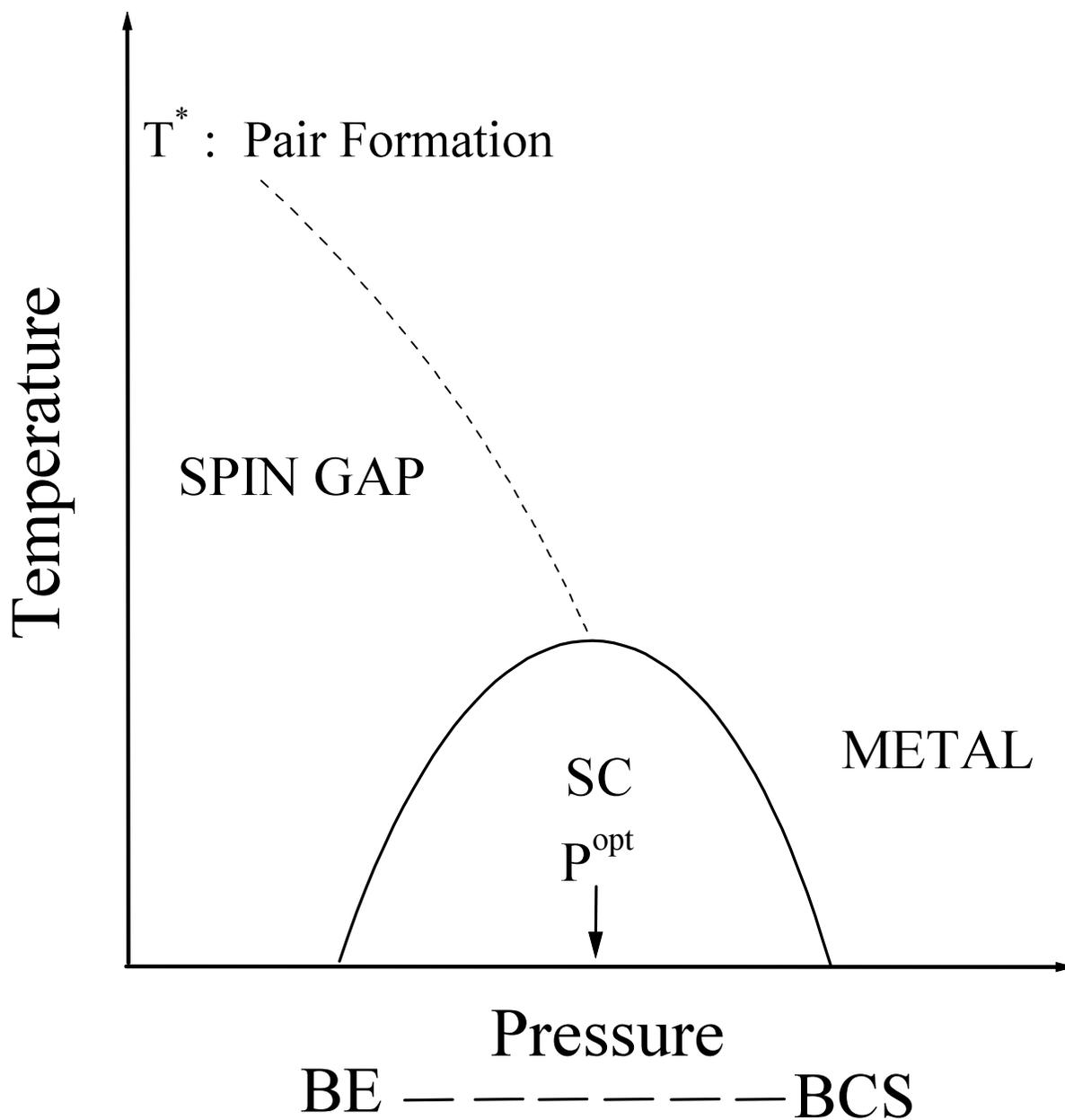}
\caption{Qualitative phase diagram of the pair formation temperature $
T^{*}$ and superconducting transition temperature $T_{c}$ as a function of the
pressure.}
\label{}
\end{figure}

\end{document}